\documentclass[twocolumn]{svjour3}                     % onecolumn (standard format)
\smartqed  % flush right qed marks, e.g. at end of proof
\usepackage{graphicx}
\usepackage{algorithmic}
\usepackage{algorithm}
\usepackage{epsfig}
\begin{document}

\title{Avoidance of Multicast Incapable Branching Nodes for Multicast Routing in WDM Networks
}
\subtitle{} %\\ Do you have a subtitle?If so, write it here


\author{Fen Zhou \and Mikl\a'os Moln\a'ar \and Bernard Cousin%etc.
}


\institute{A preliminary version of part of this work was presented in IEEE LCN 2008, Canada.\\~\\
              Fen Zhou \at
              IRISA / INSA de Rennes, Campus de Beaulieu, Rennes 35042, France   \\
              Tel.: +33-299847194\\
              Fax: +33-299847171\\
              \email{fen.zhou@irisa.fr}           %  \\
           \and
           Mikl\a'os Moln\a'ar \at
              IRISA / INSA de Rennes, Campus de Beaulieu, Rennes 35042, France
           \and
           Bernard Cousin \at
              IRISA / University of Rennes 1, Campus de Beaulieu, Rennes 35042, France
}

\date{Received: date / Accepted: date}

\maketitle

\begin{abstract}
In this paper we study the multicast routing problem in all-optical WDM networks under the spare light splitting constraint. To implement a multicast session, several light-trees may have to be used due to the limited fanouts of network nodes. Although many multicast routing algorithms have been proposed in order to reduce the total number of wavelength channels used (\emph{total cost}) for a multicast session, the maximum number of wavelengths required in one fiber link (\emph{link stress}) and the \emph{end-to-end delay} are two parameters which are not always taken into consideration. It is known that the shortest path tree results in the optimal \emph{end-to-end delay}, but it can not be employed directly for multicast routing in sparse light splitting WDM networks. Hence, we propose a novel wavelength routing algorithm which tries to avoid the multicast incapable branching nodes (MIBs, branching nodes without splitting capability) in the shortest-path-based multicast tree to diminish the \emph{link stress}. Good parts of the shortest-path-tree are retained by the algorithm to reduce the \emph{end-to-end delay}. The algorithm consists of three steps:~(1) a DijkstraPro algorithm with priority assignment and node adoption is introduced to produce a shortest path tree with up to 38\% fewer MIB nodes in the NSF topology and 46\% fewer MIB nodes in the USA Longhaul topology, (2) critical articulation and deepest branch heuristics are used to process the MIB nodes, (3) a distance based light-tree reconnection algorithm is proposed to create the multicast light-trees. Extensive simulations demonstrate the algorithm's efficiency in terms of \emph{link stress} and \emph{end-to-end delay}.
\end{abstract}

\section{Introduction}
\label{Introduction}
Multicast is a very efficient approach for one-to-many or many-to-many communication. A multicast session typically involves a source and a set of destinations. Typically in IP-based packet-switching networks a multicast tree rooted at the source is constructed with branches spanning all destinations to accommodate a single source-based multicast session. In order to be able to multicast data in WDM optical networks, optical switches need to have splitting capability. Note however that optical switches with light splitters are always far more expensive to build than those without. Consequently when only a few nodes can support splitting the network is described as sparse light splitting~\cite{rMalli1998}. Hence, multicast routing in WDM optical networks is very different from that in IP-based packet switching networks and one must consider the constraint on node splitting capability in a practical optical network. To implement multicast in all-optical WDM networks, the light-tree concept was proposed in~\cite{lhSahasrabuddhe1999}. A light-tree is a point-to-multipoint generalization of a lightpath containing one continuous lightpath from the source to each destination. To create light-trees, optical constraints must be respected. Without wavelength conversion the same wavelength must be used on all links in a light-tree. This is called the wavelength continuity constraint~\cite{bMukherjee2000}. Moreover, two light-trees or lightpaths sharing a common link must be assigned different wavelengths. This is known as the distinct-wavelength constraint~\cite{bMukherjee2000}. Due to these physical constraints, supporting multicast routing in all-optical networks is a challenging task.

For multicast routing in WDM optical networks many multicast light-tree computation algorithms have been proposed to reduce the total number of wavelength channels used (i.e., the total cost), but the maximum number of wavelengths required in one fiber link (i.e., the link stress) and the end-to-end delay are also important factors which should be taken into account. This is especially true for time sensitive and bandwidth intensive multicast applications such as HDTV, VoIP and Video Conference. It is known that if a message is transmitted via the shortest path from the source to the destination, then, in general, the delay is minimal. Unfortunately, splitting nodes are very rare in optical networks due to their high cost and complex architecture. If a shortest path tree is directly applied for multicast routing in WDM networks, then there is a high probability that the branching nodes of this tree do not coincide with multicast capable switches. When this is the case different wavelengths must be used to send messages from the source to different branches of a multicast incapable branching node, and the stress on the commonly used links will be very high. If the shortest paths are not used for all destinations, then a destination could find a longer path to the source (e.g., by connecting to a nearby splitting node), which implies a bigger end-to-end delay. Thus, a tradeoff must be found between link stress and end-to-end delay in order to obtain the best general performance.

In this paper, a multicast routing algorithm considering sparse light splitting is proposed which tries to avoid multicast incapable branching nodes in multicast light-trees. It aims to reduce both the link stress and the end-to-end delay. The significant aspects of this proposition are: (i) a DijkstraPro algorithm with priority assignment and node adoption, introduced to construct a shortest path tree with fewer multicast-incapable branching nodes, (ii) critical-articulation and deepest-branch heuristics are used to process the MIB nodes with the aim of reducing both link stress and end-to-end delay, (iii) a distance-based light-tree reconnection algorithm is proposed to create a set of multicast light-trees with smaller end-to-end delay while keeping the same link stress and total cost.

The rest of the paper is organized as follows: Related work is reviewed in Section~\ref{sec:Related Work}. The wavelength routing problem under the sparse light splitting model is formulated and some necessary definitions are given in Section~\ref{sec:Multicast Routing Under Sparse Light Splitting Constraint}. The multicast routing algorithm based on avoidance of multicast incapable branching nodes is proposed and simulated in Sections ~\ref{sec:Avoidance of MIB Nodes for Multicast Routing} and ~\ref{sec:Performance Evaluation and Simulation}. Finally, a summary of results is made in Section ~\ref{sec:Conclusion}.

\section{Related Work}
\label{sec:Related Work}
The difficulty of multicast routing in WDM networks with sparse light splitting has been addressed in many papers\cite{xjzhang2000,nSreenath2001Photonic,nSreenath2001High,aHamad2006,aZsigri2003,sgYani2003,fZhou2008LCN,fZhou2008ICCS} and various algorithms have been proposed. There are broadly three main categories according to the routing approaches they employ: \emph{Shortest Path Tree Based Routing} (e.g., Reroute-to-Source and Reroute-to-Any~\cite{xjzhang2000}), \emph{Steiner-Based Routing} (e.g., Member-Only ~\cite{xjzhang2000} and Virtual-Source Capacity-Priority ~\cite{nSreenath2001Photonic}) and \emph{Core-Based Routing} (e.g., Virtual Source-based ~\cite{nSreenath2001High,aHamad2006}). Essentially, the \emph{Shortest Path Based Routing} approach constructs the multicast tree by connecting the source to each destination individually using the appropriate shortest path in order to minimize the per-source-receiver path cost. The objective of the \emph{Steiner-Based Routing} schemes, however, is to minimize the overall cost of the multicast light-trees. The \emph{Core-Based Routing} algorithm connects a subset of nodes, called core nodes, which have both light-splitting and wavelength-conversion capacities. The multicast session is then established with the help of this core structure ~\cite{nSreenath2001High,aHamad2006}. To the best of our knowledge from the literature \cite{xjzhang2000}, in WDM networks with sparse-splitting and without wavelength conversion the Member-Only algorithm yields the approximate minimal cost and the best link stress, while the Reroute-to-Source algorithm yields the optimal delay.

In Reroute-to-Source, a multicast tree is first generated to span all destinations, for example by computing the shortest path tree with the Dijkstra algorithm. Then, it checks the light splitting capability for each branching node in the shortest path tree. If a branching node is a node with splitting capability, then no modification is needed. But if it is a multicast incapable branching node (i.e., it has at least two direct children while it has no splitting capability), then only one direct child can be kept, which is chosen arbitrarily. All other direct children (sub-trees) must be connected to the source through the shortest path, each on a different wavelength. It is obvious that the end-to-end delay of Reroute-to-Source is minimal. The stress of the link is very high, however, because downstream branches of a multicast incapable branching node must be connected to the source using the same shortest path on several different wavelengths. Note that there may actually be some longer paths leading to the source available on the same wavelength that are not used.

In Reroute-to-Any, similarly to Reroute-to-source, a shortest path tree is first computed for all the destinations, and for each multicast-incapable branching node one downstream branch is kept and the others are cut. Finally, the cut destinations are reconnected to the multicast light-tree via a multicast-capable node or a leaf multicast-incapable node in the light-tree if possible. If this is not possible they are reconnected to the source on different wavelengths. Although its link stress and total cost are better than the Reroute-to-Source and its average end-to-end delay is superior to Member-Only, the algorithm is still not entirely satisfactory and should be improved to take traffic with QoS requirements into account. It seems that no algorithm has been proposed to decide which branch of the multicast-incapable branching nodes should be kept and what kind of reconnection algorithm can be used to reconnect the cut destinations.

In Member-Only, each iteration adds the nearest destination to the multicast tree using the shortest path. The shortest path must not include any non-leaf multicast incapable nodes in the light-tree under construction. It is a modification of the well known Takahashi-Matsuyama heuristic of the Steiner problem~\cite{hTakahashi1980,aZsigri2003}. Although its total cost approaches the optimum, it is very likely that most of the destinations are connected to the source via a node far away from the source. As a result, its average end-to-end delay is high and the diameter of the multicast tree is always very large.

\section{Multicast Routing Under Sparse Light Splitting Constraint}
\label{sec:Multicast Routing Under Sparse Light Splitting Constraint}
\subsection{System Model and Problem Formulation}
\label{sec:System Model}
One-to-many communication in all-optical WDM networks is now examined. Network nodes equipped with light splitters are assumed to be sparse because of their complex architecture and expensive cost, with a presence normally below 50\% ~\cite{mAli2000Allocation}. Furthermore, it is assumed that costly wavelength converters are not available. With no loss of generality, the splitting capability of a multicast capable node is assumed to be infinite by supposing correct use of optical amplifiers~\cite{eDesurvire1991}. A spare light splitting WDM network can be modeled by an undirected graph $G (V, E, c, d)$. Each node $v \in V$ is either a multicast incapable node (without splitting capability) or a multicast capable node (equipped with light splitters). Each edge $e \in E$ is associated with two weight functions $c(e)$, $d(e)$. $c(e)$ represents the cost of fiber link $e$, and $d(e)$ denotes the propagation delay in fiber link $e$. Both of them are additive along a lightpath $LP(u, v)$. We consider the arrival of a multicast session $ms(s, D)$, which requires a simultaneous communication from the source $s$ to be set up to a group of destinations $D$. Due to the sparse-splitting constraint together with the wavelength-continuity constraint, one light-tree may not be sufficient to cover all destinations. Assume $k$ light-trees $LT_{i}(s, D_{i})$ should be built for a multicast session $ms(s, D)$ where $i \in [1, k]$ and $D_{i}$ denotes the set of destinations exclusively served in the $i^{th}$ light-tree. Since these $k$ light-trees are not edge disjoint, different wavelengths must be assigned for each light-tree. Thus, the number of wavelengths required for $ms(s,D)$ (i.e., link stress) equals the number of light-trees built.

  \begin{equation}
    Stress[ms(s,D)]=k
  \end{equation}

  The total number of wavelength channels used (i.e., total cost) for $ms(s,D)$ can be calculated as
  \begin{equation}
    c[ms(s,D)]=\sum_{i \in [1,k]} \sum_{e \in LT_{i}(s,D_{i})} {c(e)}
  \end{equation}

Nowadays multimedia services such as HDTV, VoIP, Video Conference and Video on Demand are widespread in the Internet. They are delay sensitive and bandwidth intensive. Consequently, the link stress and the delay are two important parameters for multicast light-tree selection in WDM optical networks. If the link stress is very high for one multicast session, fewer wavelengths can be allocated for other sessions. This may lead to a reduction of network throughput.

Besides this, power loss and noise are two other challenging problems in all-optical networks. Although power loss can be compensated by appropriate placement of optical amplifiers in fibers and cross-connects, noise resulting from amplification can cascade and is hard to remove without electronic processing. It is practical to limit the length of a path in order to decrease the number of amplifiers~\cite{sgYani2003}. In addition, optical networks are becoming increasingly widespread in the Internet Backbone. Although optical messages are transmitted from the source to the destination at a very high speed, the nodes in WDM optical networks can be distributed around the world. When this is the case, the end-to-end delay will not be negligible and delay-sensitive traffic will require special treatment. Let $LP(s,d_{i})$ be the lightpath between the source $s$ and the destination $d_{i}$ in the light-trees built for a multicast session $m(s,D)$, the average end-to-end delay and the maximum end-to-end delay can be defined as follow:

  \begin{equation}
    AverDelay[ms(s,D)]=\frac{1}{|D|} \sum_{d_{i} \in D} \sum_{e \in LP(s,d_{i})} {d(e)}
  \end{equation}

  \begin{equation}
    MaxDelay[ms(s,D)]=\max_{d_{i} \in D} \sum_{e \in LP(s,d_{i})} {d(e)}
  \end{equation}

Note that end-to-end delay and link stress cannot be minimized simultaneously. If the shortest path tree is directly used for multicast routing, although its delay is optimal, its link stress is generally very high. When an approximated Sterner tree is employed to build the multicast light-trees (using the Member-Only algorithm ~\cite{xjzhang2000}), the link stress is good, but the end-to-end delay is intolerable: an approach that produces a tradeoff solution needs to be found. In order to minimize the end-to-end delay, the shortest path tree can be considered a good starting point for the construction of multicast light-trees. In order to improve the link stress, the number of MIB nodes in the shortest path tree can be reduced by making some destinations communicate with the source using longer paths. Putting this approach into practice, a multicast routing algorithm based on avoidance of MIB nodes is proposed for WDM networks with sparse light splitting. To simplify the objective metrics, the same edge cost and delay functions are applied throughout the network, and without loss of generality they are given as:

  \begin{equation}
    c(e) = 1~unit~cost \\
    d(e) = 1~unit~delay\\
  \end{equation}

\subsection{Useful Definitions}
\label{sec:Useful Definitions}
Before describing our proposed multicast routing algorithm, some necessary definitions are introduced below.
\begin{description}
  \item Definition 1:  MI and MC nodes

   MI nodes: multicast incapable nodes are nodes which cannot split light signals, but have a \emph{Tap-and-Continue} (\emph{TaC}~\cite{mAli2000Cost}) capability. That is to say, they can tap a small amount of optical power from a wavelength channel while forwarding it to only one output link.

%DG check a/an MI

   MC nodes:  multicast capable nodes are nodes which are equipped with light splitters which can distribute the incoming message to all of the outgoing ports.\\

   In the illustration figures (except those for network topology, Figures~\ref{NSFnet} and \ref{longhual}) MI nodes are denoted by a rectangle while MC nodes are denoted by circles.\\

  \item Definition 2: Multicast Incapable Branching Node (MIB node)

    MIB nodes have no splitting capability, but lead to several downstream branches in a multicast tree. An MIB node's out degree in the multicast tree is not less than two. Once an MIB node has forwarded the message to one branch it is incapable of forwarding it to another branch using the same wavelength.\\

    \begin{figure}
        \begin{center}
        \includegraphics[width=.35\textwidth]{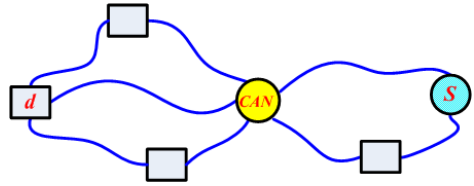}
        \end{center}
        \caption{Critical Articulation Node}
        \label{An}       % Give a unique label
   \end{figure}

  \item Definition 3:  Set MC{\_}SET, MI\_SET and D

   A set of light-trees may be required by a multicast session. For a multicast light-tree $T$ under construction,

   MC\_SET: includes the MC nodes and the leaf MI nodes in $T$. They may be used to span $T$. Hence, nodes in MC\_SET are also called connector nodes in $T$.

   MI\_SET: includes only non-leaf MI nodes in $T$ which are unable to connect a new destination to $T$.

   D: includes unvisited multicast members which are neither joined to the current multicast light-tree $T$ nor to the previously constructed multicast light-trees.\\

  \item Definition 4: Constrained Path (CP) and Shortest Constrained Path (SCP)

   A Constrained Path $CP(u,T)$ between a node $u$ and a tree $T$ is defined as the shortest path $SP(u,v)$ from node $u$ to a connector node $v$ in $T$, such that $SP(u,v)$ does not traverse any node belonging to the MI\_SET of $T$.

   \begin{equation}
         \begin{array}{ccll}
         CP(u,T) & = & \{SP(u,v)|v \in \emph{MC\_SET}, and &~~~~~~~\\
            &  &\multicolumn{2}{r}{\forall x \in SP(u,v), x\not \in \emph{MI\_SET}\}}
         \end{array}
   \end{equation}

   Accordingly, the shortest of all possible Constrained Paths $CP\_Set(u,T)$ is called the Shortest Constraint Path $SCP(u,T)$.

  \begin{equation}
    c[SCP(u,T)] = \min_{CP(u,T) \in CP\_Set(u,T)}{c[CP(u,T)]}
  \end{equation}
  There may be several $SCP(u,T)$ from $u$ to $T$ with different connector nodes $v$.\\
   % \begin{equation}
%    CP(u,T) = \{p(u,v)|v \in \emph{MC\_SET}, and \ \forall x \in p(u,v), x\not \in \emph{MI\_SET}\} \raggedleft
%    \end{equation}

  % \begin{equation}
%   CP(u,T)=\left \{
%     \begin{array}{ll}
%         p(u,v)|v \in \emph{MC\_SET}, and &~~~~~~~~~~\\
%         \multicolumn{2}{r}{\forall x \in p(u,v), x\not \in \emph{MI\_SET}}
%         \end{array}
%   \right\}
%      \end{equation}

  \item Definition 5: Connection Constraint Node (CC node) and Critical Articulation Node (CAN)

   If node $u$ is a CC node, there must be an intermediate node which is included in all the paths from $u$ to the source $s$. This intermediate node is called the critical articulation node: $CAN(u, s)$. In other words, a CC node $u$ cannot reach the source $s$ without node $CAN(u, s)$.

   \begin{algorithm}
    \algsetup{indent=2em}
    \caption{~~~Avoidance of MIB Nodes for Multicast Routing}
    \label{alg0}
    \begin{algorithmic}[1]
    \REQUIRE {A multicast session $ms(s,D)$}
    \ENSURE {A set of multicast light-trees for $ms(s,D)$}
    \STATE Use the DijkstraPro algorithm to construct the shortest path tree SPT rooted at the source $s$. Prune all the non-destination leaf nodes and the nodes which do not lead to any destinations.
    \STATE Use the Deepest Branch and the Critical Articulation Heuristics to process the MIB nodes in SPT.
    \STATE Use the distance based light-trees reconnection algorithm to create the required set of light-trees for $ms(s,D)$.
    \end{algorithmic}
    \end{algorithm}

   For example, in Figure~\ref{An}, node $CAN$ separates the network into two parts. Node $d$ and source $s$ are in different parts. Without node $CAN$, $d$ is not able to communicate with $s$. So $d$ is a CC node, and node $CAN$ is the $CAN(d, s)$.
   \end{description}

\section{Avoidance of MIB Nodes for Multicast Routing}
\label{sec:Avoidance of MIB Nodes for Multicast Routing}
The avoidance of MIB nodes multicast routing algorithm can be viewed as a post-processing of the shortest path tree (SPT). Due to the presence of MIB nodes in a shortest path tree a single wavelength may not be sufficient to cover all destinations and thus several wavelengths may be required to accommodate the multicast group. Thus, MIB nodes should be avoided in order to decrease the link stress. If there are no MIB nodes in the shortest path tree, then the tree is an optimal multicast light-tree with both minimum end-to-end delay and minimum link stress. If this is not the case some processing must be done on the MIB nodes. The proposed algorithm consists of three main steps: the shortest path tree construction step, the MIB nodes processing step and the multicast light-tree reconstruction step. In the first step, an enhanced DijkstraPro algorithm making use of the priority method and node adoption is introduced to construct a multicast shortest path tree with fewer MIB nodes and smaller link stress. In the second step the MIB nodes in the shortest path tree are processed: deepest branch and critical articulation heuristics are proposed to keep only one downstream branch of MIB nodes in an attempt to reduce both the link stress and the end-to-end delay. In the last step the distance-based light-tree reconnection algorithm (which can also reduce end-to-end delay) is applied to create the multicast light-trees.

\subsection{Construction of SPT and DijkstraPro Algorithm}
\label{sec:Construction of  SPT and DijkstraPro Algorithm}
First of all, a shortest path tree rooted at the source is constructed for all nodes in the network. Then, according to the multicast session, non-destination nodes and nodes that do not lead to any destination are pruned from this shortest path tree.

Generally, Dijkstra's algorithm is employed to build the shortest path tree. In the Dijkstra algorithm, a node is said to be labeled permanently~\cite{csrMurthy2002} if its shortest path to the source has been found. Otherwise it is said to be tentatively labeled~\cite{csrMurthy2002}. Initially, only the source $s$ is permanently labeled and all the other nodes are tentatively labeled. In each iteration, the node with the shortest distance to the source among all the tentatively labeled nodes is chosen and labeled permanently. It is worth noting that in one iteration there may be several nodes that have the same shortest distance to the source, here we call them \emph{candidate nodes} and the distance is referred to as their \emph{level}. However, according to the Dijkstra algorithm we should label only one of the candidate nodes permanently in order to update the distances of the other nodes. The question then, is how to choose the permanently labelled candidate node? In the traditional Dijkstra algorithm, it is chosen arbitrarily. But consider this situation: there are two candidate nodes at the same \emph{level}; one is a MI node and another is a MC node; they share the same two adjacent nodes. If the MI \emph{candidate} node is the first to be selected for permanent labelling then the two adjacent nodes will update their distances to the source, and thus will be connected to the source via this MI \emph{candidate} node. The problem is that the MI \emph{candidate} node cannot split the incoming signal to more than one outgoing port. As a result, it will become a MIB node in the shortest path tree. Alternatively, if the MC \emph{candidate} node is the first to be permanently labeled then when the two adjacent nodes update their distances to the source they will be connected to it via this MC \emph{candidate} node. Subsequently, the MI candidate node is chosen to be permanently labeled. At this point, no adjacent node needs to update its distance and no adjacent node is left to be connected to the source via this MI \emph{candidate} node. So, the risk that an MI \emph{candidate} node will become an MIB node is reduced or even avoided.

Due to the constraint on splitting capability, the traditional Dijkstra algorithm may not yield a favorable result, but it can be improved with some modification. Hence, the DijkstraPro algorithm with priority and node adoption is presented. When building a shortest path tree using Dijkstra with several \emph{candidate} nodes at the same \emph{level} the following operations are proposed:

    \begin{itemize}
    \item \emph{Giving Higher Priority to MC Candidate Node} \\
    The \emph{candidate} node with multicast splitting capability (MC \emph{candidate} node) should be given higher priority than the MI \emph{candidate} nodes due to the fact that they can connect many destination nodes to the tree without producing a MIB node.  In other words, the probability that an MI \emph{candidate} node will be used to connect more than one destination to the tree in latter iterations is greatly reduced.

    Refer to the NSF network in Figure~\ref{NSFnet}. Nodes 1, 8 and 10 are assumed to be MC nodes. A multicast session arrives: \emph{m$_{1}$}= \{source: 10 $\mid$ members: 1 $\sim$ 14\}. If the Dijkstra algorithm is used then we can get the shortest path tree in Figure~\ref{spt1}. There are 2 MIB nodes in this shortest path tree. We can see, however, that nodes 1, 6, 7, 9 and 13 have the same shortest distance to the source node 10. So, they can be viewed as \emph{candidate} nodes at the same \emph{level}. And, if node 1 (an MC node) is promoted to a higher priority and chosen first to be permanently labeled, followed by 7, 9, 13 and 6, then we can get the new shortest path tree of Figure~\ref{spt2} which has only one MIB node.\\

    \item \emph{Giving High Priority to MI Candidate Nodes with Smaller Degree}\\
    If there are no MC \emph{candidate} nodes, then the \emph{candidate} node with the smaller degree is given a higher priority. This is because the probability that an MI \emph{candidate} node with a smaller degree is a MIB node is very low (especially for \emph{candidate} nodes with a degree of two). That is to say, the number of nodes that remain to be connected to the source through other MI \emph{candidate} nodes with higher degrees is very small. Consequently, the probability for a \emph{candidate} node with a higher degree to become an MIB node is reduced. So, the average probability for a node to be a MIB node is slightly reduced. \\
	
    \item \emph{Node Adoption}\\
    At the stage when all \emph{candidate} nodes at the same \emph{level} have been permanently labeled, the following situation may occur: some MI \emph{candidate} nodes connect only two direct children to the tree (i.e., MIB \emph{candidate} nodes) while some \emph{candidate} nodes are leaf nodes in the created tree. Thus, the possibility arises for a leaf \emph{candidate} node to adopt one child from an MIB \emph{candidate} node at the same \emph{level} when the child can reach the source through the leaf \emph{candidate} node also. By doing this the creation of an MIB node can be avoided. Node adoption between the \emph{candidate} nodes at the same \emph{level} can result in a greatly reduced number of MIB nodes in the shortest path tree or in the balancing of the load of an MIB node. Typically a destination node should be given a higher priority when determining which nodes may be adopted.

  \begin{figure}
        \begin{center}
        \includegraphics[width=.4\textwidth]{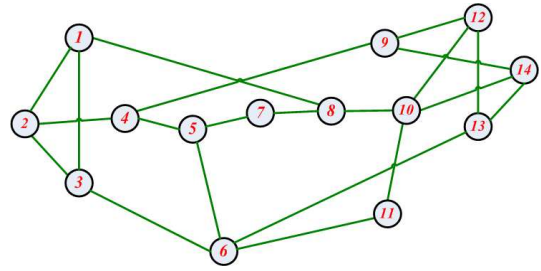}
        \end{center}
        \caption{NSF Network Topology}
        \label{NSFnet}       % Give a unique label
    \end{figure}

    \begin{figure}
           \begin{center}
            \includegraphics[width=.25\textwidth]{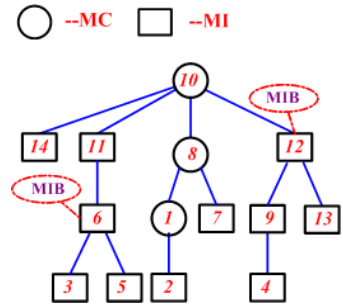}
           \end{center}
            \caption{The SPT for \emph{m$_{1}$} constructed by the Dijkstra algorithm}
            \label{spt1}       % Give a unique label
    \end{figure}

    \begin{figure}
           \begin{center}
            \includegraphics[width=.25\textwidth]{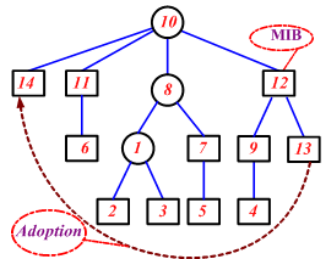}
           \end{center}
            \caption{The SPT for \emph{m$_{1}$} constructed by offering higher priority to MC \emph{candidate} nodes}
            \label{spt2}       % Give a unique label
    \end{figure}

    \begin{figure}
            \begin{center}
            \includegraphics[width=.25\textwidth]{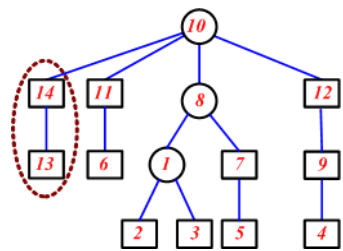}
            \end{center}
            \caption{The SPT after Node Adoption from Figure~\ref{spt2}}
            \label{spt3}       % Give a unique label
    \end{figure}

    Refer to the example in Figure~\ref{spt2}. It is obvious that nodes 11, 12 and 14 have the same least distance to the source node 10, hence they can be viewed as \emph{candidate} nodes at the same \emph{level}. After all of them have been permanently labeled we can see that node 12 is an MIB node and node 14 is a leaf node. Note that nodes 13 or 9 can reach the source node 10 by the shortest path through both of nodes 12 and 14. Thus, one of them could be adopted by node 14, and a new shortest path tree without an MIB node can be obtained as in Figure~\ref{spt3}.
  \end{itemize}

    \subsection{Processing of the MIB nodes}
    \label{sec:Processing of the MIB nodes}
    Due to the fact that the MIB nodes in the shortest path tree can forward the incoming message to one and only one outgoing branch, the existence of MIB nodes is the most important cause of high link stress. Thus, they should be processed and avoided. In the Reroute-to-Source algorithm~\cite{xjzhang2000}, all downstream links of MIB nodes are connected to the source through the reverse shortest path on different wavelengths which results in high link stress. Although the Reroute-to-Any algorithm is also proposed in the literature~\cite{xjzhang2000}, there is no description of how to keep one branch when processing the MIB nodes. So, in this paper, the deepest branch and critical articulation heuristics are employed to decide which branch should be kept in order to decrease the link stress and the end-to-end delay.

  \subsubsection{MIBPro}
   \label{sec:MIBPro}
   \begin{itemize}
     \item \emph{Critical Articulation Heuristic} \\
        A CC node $u$ can only communicate with the source through its $CAN(u, s)$. In a multicast tree, if the $CAN(u, s)$ is (unfortunately) an MIB node, then the branch containing $u$ should be assigned a higher priority and kept when processing this MIB node. This is because there is no alternative path for $u$ to reach the source without traversing its $CAN(u, s)$. However, destinations in the other branches may find another path to the source which will not traverse this MIB node. In fact, CC and $CAN(CC, s)$ nodes are very rare in real optical networks. However, in the case that some nodes in the network have failed they may exist, and this heuristic will be very practical. In the network of Figure~\ref{examplenet}, node $d_{1}$ is a CC node. The shortest path tree for multicast session $m_{2}$ = \{source: s $\mid$ destinations: $d_{1} \sim d_{6}$ \} is given by Figure~\ref{examplenet}. We can see that $CAN(d_{1}, s)$ is an MIB node in the shortest path tree built for \emph{m$_{2}$} as plotted in Figure~\ref{spt-noCA}, hence it should be processed. If node $d_{1}$ is disconnected from $CAN(d_{1}, s)$ and the branch leading to node $d_{2}$ and $d_{3}$ is kept, then two light-trees on two different wavelengths $w_{0}$ and $w_{1}$ are required as shown in Figure~\ref{spt-noCA}. But if the CC node $d_{1}$ is kept and the other one is cut, then only one light-tree (\emph{or one wavelength}) is needed as shown in Figure~\ref{spt-CA}.\\

        \begin{algorithm}
    \algsetup{indent=2em}
    \caption{~~~Processing of MIB nodes Using Critical Articulation and Deepest Branch Heuristics}
    \label{alg1}
    \begin{algorithmic}[1]
    \STATE Search all the MIB nodes in the shortest path tree
    \FOR {each MIB node}
      \IF{No downstream branch contains a CC node}
      \STATE Keep the deepest branch
      \ELSIF{Only one downstream branch contains a CC node \&  MIB node  =  $CAN(CC, s)$}
      \STATE Keep the branch with the CC node
      \ELSIF{Several downstream branches contain CC nodes \&  MIB node  =  $CAN(CC_{i}, s), i =1, 2,\ldots$}
      \STATE Keep the deepest branch with a CC node
      \ENDIF
    \ENDFOR
    \STATE Delete the downstream branches of MIB nodes which are not kept
    \end{algorithmic}
    \end{algorithm}

    \begin{figure}
            \begin{center}
            \includegraphics[width=.28\textwidth]{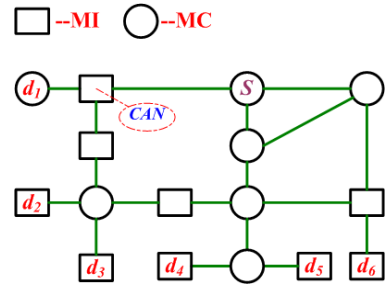}
            \end{center}
            \caption{An example network with CAN nodes}
            \label{examplenet}       % Give a unique label
    \end{figure}

    \begin{figure}
            \begin{center}
            \includegraphics[width=.25\textwidth]{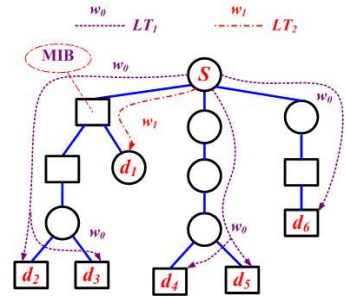}
            \end{center}
            \caption{A Shortest Path Tree for \emph{m$_{2}$}}
            \label{spt-noCA}       % Give a unique label
    \end{figure}

    \begin{figure}[h!]
            \begin{center}
            \includegraphics[width=.22\textwidth]{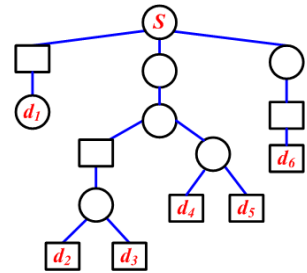}
            \end{center}
            \caption{Processing MIB nodes using the Critical Articulation Heuristic}
            \label{spt-CA}       % Give a unique label
    \end{figure}

     \item \emph{Deepest Branch Heuristic} \\
         The deepest branch should also be assigned a higher priority. This is because a destination far away from the source has difficulty finding a path to the source without traversing a non-leaf MI node in the tree. Furthermore it is desirable to minimise the average end-to-end delay for a destination node far away from the source by choosing the shortest path to the source. To implement this step a breadth-first traversal algorithm can be employed. The worst case time complexity of this heuristic is $O(N)$, where $N=|V|$ is the number of nodes in the network.
   \end{itemize}

   \subsubsection{MIBPro2}
   \label{sec:MIBPro2}
   In addition another method is proposed to process MIB nodes in the shortest path tree. This method deletes all the downstream branches of an MIB nodes without employing any heuristic. These two methods will be compared in Section~\ref{sec:Performance Evaluation and Simulation}.

   \subsection{Reconnection of Multicast Light-trees}
   \label{sec:Reconnection of Multicast Light-trees}
   After the MIB node processing step, the shortest path tree is divided into a disconnected forest containing a subtree plus several separated destinations. This disconnected forest must be reconnected in order to accommodate all the multicast members. A Member-Only-like \cite{xjzhang2000} light-tree connection method would be a good candidate to reconnect the multicast forest. The Member-Only algorithm always adds the destination nearest to the multicast light-tree using the shortest path, but this shortest path will not use any non-leaf MI node in the light-tree. In other words, at each iteration only the destination with the shortest SCP is connected to the light-tree using this SCP. As demonstrated in~\cite{xjzhang2000}, the Member-Only algorithm can achieve the best link stress and the minimum cost, although its end-to-end delay is very large. It is worth noting that some improvements can be made to this algorithm to reduce the end-to-end delay to some extent while obtaining the same cost and the same link stress. The example below demonstrates how end-to-end delay can be reduced.

    \begin{algorithm}
    \algsetup{indent=2em}
    \caption{~~~Distance Based Light-tree Reconnection Algorithm}
    \label{alg1}
    \begin{algorithmic}[1]
    \STATE {$T \leftarrow \emph{\textbf{subtree}}~$obtained~after~MIB~process$ $}
    \STATE {$MC\_SET \leftarrow \{MC~nodes~$and$~leaf~MI~$nodes~in$~T\}$}
    \STATE {$MI\_SET \leftarrow \{non$-$leaf~MI~$nodes~in$~T\}$}
    \STATE {$D \leftarrow \{destinations~$not~in$~T\}$}
    \WHILE{($D \neq \Phi$)}
        \REPEAT
        \STATE Find the closest destination $d \in D$  to \emph{T}, such that its shortest path
               to \emph{T} does not traverse any node in \emph{MI\_SET}
        \IF {there are several destinations satisfying equation~\ref{equ1}}
        \STATE Select the destination nearest to $s$ in network \emph{G} as \textbf{d}
        \ENDIF
        \IF {there are several connector nodes for \textbf{d} in \emph{MC\_SET} satisfing equation~\ref{equ2}}
        \STATE Select the connector node nearest to $s$ in \emph{T} as \textbf{c}
               and choose the corresponding SCP
        \ENDIF
        \STATE  $T \leftarrow T \cup SCP(\textbf{d},\textbf{c})$
        \STATE  $MC\_SET \leftarrow  MC\_SET \cup \{\textbf{d}~$and$~MC\-nodes~$on$~SP(\textbf{d}, \textbf{c})\}$
        \STATE  $MI\_SET \leftarrow  MI\_SET \cup \{non$-$leaf~MI\-nodes~$on$~SP(\textbf{d}, \textbf{c})\}$
        \STATE  $D \leftarrow D \setminus d$
        \IF {\textbf{c} is an \emph{MI} node}
            \STATE $MC\_SET \leftarrow MC\_SET \setminus \textbf{c}$
            \STATE $MI\_SET \leftarrow MI\_SET \cup \{\textbf{c}\}$
        \ENDIF
        \UNTIL {no destination can be added to \emph{T}}
        \RETURN {\emph{T}}
        \STATE $ $Begin~a~new~tree$~T \leftarrow \{s\}$
        \STATE $MC\_SET \leftarrow \{s\}$
        \STATE $MI\_SET \leftarrow \phi $
    \ENDWHILE
    \algsetup{}
     \begin{equation}\label{equ1}
        dist\{SCP(d, T )\} = \min_{d_{i} \in D} {dist[SCP(d_{i}, T)]}%
     \end{equation}
     \begin{equation}\label{equ2}
        dist\{SCP(\textbf{d}, T)\} = dist\{SP(d, connector_{i})\}, i=1,2,\ldots
     \end{equation}
    \end{algorithmic}
    \end{algorithm}

\begin{figure}[!t]
            \begin{center}
            \includegraphics[width=.32\textwidth]{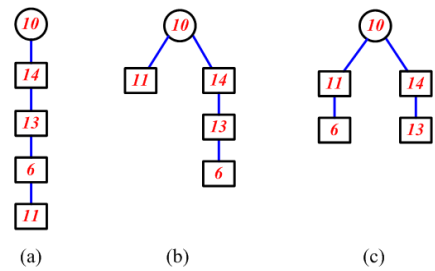}
            \end{center}
            \caption{Two strategies for the reconnection of light-trees}
            \label{twohints}       % Give a unique label
    \end{figure}

   A multicast session $m_{3}$ = \{source: 10 $\mid$ destinations: 6, 11, 13, 14\} is required in the NSF network, Figure~\ref{NSFnet}. We assume that the first tree only contains the source node 10. According to the previously described member-only-like light-tree reconnection approach, the destination with the shortest SCP should be added to this tree first. The shortest paths for node 11 and node 14 to the source have length 1. Without loss of generality, let us suppose node 14 is the first to be connected. Then, on the new tree, we can see that both SCPs for nodes 11 and 13 have the same length. Also without loss of generality, suppose node 13 is then connected. After that node 6 is chosen, and finally node 11. Following these steps, the resultant multicast tree is given in Figure~\ref{twohints}(a).

   It is immediately apparent that node 11 can be connected to the tree via node 10 or node 6.  Why do we not connect it through node 10 as in Figure~\ref{twohints}(b)? The difference is that the connector nodes have different distances to the source in the tree (for node 10 the distance is 0 while that for node 6 is 3). In addition, it is even more interesting to consider Figure~\ref{twohints}(c). All three of these multicast trees have the same cost of 4 while having different average delays: 10/4, 7/4 and 6/4. It is also simple to determine that following the addition of node 14 to the tree, if node 11 is added before node 13 we can get the result in Figure~\ref{twohints}(c).
	
   So, from this simple example we have two strategies that reduce the average delay while maintaining the same cost and the same link stress. The distance based reconnection algorithm is developed from these observations. If there are several nodes whose SCPs to the multicast tree have the same length, then these nodes should be added in the order of their distance to the source (the distance in the network): the nearer, the earlier. When the destination with the shortest SCP has at least two connector nodes in the subtree, it is better to use the connector node nearest to the source (the distance in the multicast light-tree under construction), otherwise its end-to-end delay will be too large.

%DG - the distance in the multicast...

   \section{Performance Evaluation and Simulation}
   \label{sec:Performance Evaluation and Simulation}
\begin{figure}
            \begin{center}
            \includegraphics[width=.5\textwidth]{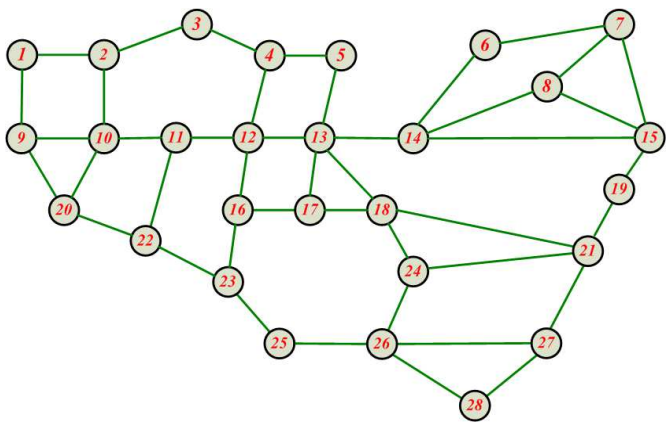}
            \end{center}
            \caption{USA Longhaul Network}
            \label{longhual}       % Give a unique label
    \end{figure}

   To ensure the effectiveness of our proposed multicast routing algorithm, two different network topologies are employed as test beds for the simulation: the 14 node NSF network in Figure~\ref{NSFnet} and the 28 node USA Longhaul network in Figure~\ref{longhual}. The fact that these networks have been used as reference topologies in many papers~\cite{nSreenath2001Photonic,aHamad2006,sgYani2003,fZhou2008LCN,fZhou2008ICCS,zgZhang2007} is the reason for their selection.

    \subsection{Performance of the DijkstraPro Algorithm}
    \label{sec:Performance of DijkstraPro Algorithm}
    To demonstrate the superiority of the DijkstraPro algorithm it is compared with the traditional Dijkstra algorithm using the following two parameters:
    \begin{itemize}
    \item N: the number of the MIB nodes in the shortest path tree.
    \item S: the maximum number of wavelengths required in one fiber link to cover all destinations in the shortest path tree (i.e., link stress of the SPT).
    \end{itemize}
	
    In each comparison, two conditions are considered. Condition 1 only regards the source to be an MC node, while Condition 2 regards nodes with a high degree to be MC nodes. The reason for choosing these two conditions can be explained as follows. In Condition 1, as only the source is an MC node, MC \emph{candidate} node priority is not applied. Thus, the result in Condition 1 checks the merit of the node adoption operation in the DijkstraPro algorithm. As stated in~\cite{hLin2005,aBillah2007}, one approach could be to place splitters at the nodes with high degree. Thus, nodes with high degree are treated as MC nodes in Condition 2. In this condition, MC \emph{candidate} node priority is applied, and the overall performance can be verified.

    \begin{table}
    \caption{Comparision of Dijkstra and DijkstraPro algorithms in the NSF Network in Figure~\ref{NSFnet}}
    \label{table1}       % Give a unique label
    \begin{tabular}{|c|c|c|c|c|}
    \hline\hline
    SPT   &   \multicolumn{2}{|c|}{ Condition1}   & \multicolumn{2}{|c|}{ Condition2}\\
    in    &   \multicolumn{2}{|c|}{ MC:  source}  & \multicolumn{2}{|c|}{ MC:  6,10, and source}\\
    NSF   &	  \multicolumn{2}{|c|}{Members: 1 $\sim$ 14} & \multicolumn{2}{|c|}{Members: 1 $\sim$ 14}\\[0.5ex]
    \hline
    \scriptsize{Source} & \scriptsize{Dijkstra} & \scriptsize{DijkstraPro} & \scriptsize{Dijkstra} & \scriptsize{DijkstraPro}\\
	ID   &	N ~~~S    & N  ~~~S  & N  ~~~S  & N  ~~~S\\ [0.5ex]
    \hline
    1    &	3~~~~4 &	3~~~~3  &	1~~~~2  &	1~~~~2 \\
    2    &	4~~~~3 &	3~~~~3  &	3~~~~3  &	2~~~~3 \\
    3    &	3~~~~4 &	2~~~~4  &	2~~~~2  &	1~~~~2 \\
    4    &	4~~~~3 &	3~~~~2  &	4~~~~3  &	3~~~~2 \\
    5    &	4~~~~4 &	2~~~~3  &	3~~~~2  &	1~~~~2 \\
    6    &	3~~~~2 &	3~~~~2  &	3~~~~2  &	3~~~~2 \\
    7    &	5~~~~5 &	4~~~~4  &	3~~~~3  &	2~~~~2 \\
    8    &	4~~~~4 &	2~~~~3  &	3~~~~2  &	1~~~~2 \\
    9    &	4~~~~3 &	3~~~~3  &	3~~~~3  &	2~~~~3 \\
    10   &	3~~~~2 &	1~~~~2  &	2~~~~2  &	1~~~~2 \\
    11   &	3~~~~4 &	3~~~~4  &	1~~~~2  &	1~~~~2 \\
    12   &	4~~~~3 &	4~~~~3  &	2~~~~2  &	1~~~~2 \\
    13   &	3~~~~4 &	2~~~~4  &	2~~~~2  &	1~~~~2 \\
    14   &	4~~~~3 &	4~~~~3  &	2~~~~2  &	1~~~~2 \\
    \hline
   $Average$  & 3.64   3.43 &	2.79 3.07 & 2.43   2.29	&  1.5 2.14\\
    \hline\hline
    \end{tabular}
    \end{table}

    In Table~\ref{table1}, we evaluate the performance of 14 shortest path trees rooted at each node of the NSF network. Source ID denotes the root of the shortest path tree built. Two conditions are considered:
    \begin{description}
      \item Condition 1 (only the source is an MC node)

      The average number of MIB nodes in the shortest path tree constructed by the DijkstraPro algorithm is 0.85 less (23\%) than when applying the traditional Dijkstra algorithm and the link stress is 0.36 smaller. This result confirms that the node adoption operation in the DijkstraPro algorithm is effective.
      \item Condition 2 (nodes 6, 10 and the source are MC nodes)

      In the NSF network, node 6 and node 10 have a high degree of 4), so they can be assumed to be MC nodes which are very useful for multicast sessions. The DijkstraPro algorithm produces a shortest path tree with fewer MIB nodes and smaller link stress for this condition also. The average number of MIB nodes is 0.93 less (38\%) and the link stress is 0.15 smaller.
    \end{description}

    In Table~\ref{table2} we also provide the performance of 28 shortest path trees rooted at each node in the USA Longhaul network.
    \begin{description}
      \item Condition 1 (only the source is an MC node)

       The DijkstraPro algorithm results in 1.75 (29\%) fewer MIB nodes on average than the traditional Dijkstra algorithm, and the link stress of the shortest path tree built by DijkstraPro is 1.64 smaller. This signifies that the effectiveness of the node adoption operation is independent of network topology.
      \item Condition 2 (nodes 10, 12$\sim$15, 18, 21, 26 and the source are MC nodes)

       In the USA Longhaul network, nodes 10, 12$\sim$15, 18, 21 and 26 have a degree equal to or above 4, so they are regarded as the MC nodes in this condition. the DijkstraPro algorithm can also produce a shortest path tree with fewer MIB nodes and smaller link stress. The average number of MIB nodes is 0.78 less (46\%) and the average link stress is 0.43 smaller.
    \end{description}

    \begin{table}
    \caption{Comparision of Dijkstra and DijkstraPro algorithms in the USA Longhual Network of Figure~\ref{longhual}}
    \label{table2}       % Give a unique label
    \begin{tabular}{|c|c|c|c|c|}
    \hline\hline
    SPT   &   \multicolumn{2}{|c|}{ Condition1}   & \multicolumn{2}{|c|}{ Condition2}\\
    in    &   \multicolumn{2}{|c|}{ MC:  source}  & \multicolumn{2}{|c|}{ MC:  10,12 $\sim$ 15, 18, }\\ Longhaul &  \multicolumn{2}{|c|}{Members: 1 $\sim$ 28} & \multicolumn{2}{|c|}{21, 26 and source}\\
           &  \multicolumn{2}{|c|}{}  & \multicolumn{2}{|c|}{ Members: 1 $\sim$ 28}\\[0.5ex]
    \hline
    \scriptsize{Source} & \scriptsize{Dijkstra} & \scriptsize{DijkstraPro} & \scriptsize{Dijkstra} & \scriptsize{DijkstraPro}\\
	ID   &	N ~~~S    & N  ~~~S  & N  ~~~S  & N  ~~~S\\ [0.5ex]
    \hline
    1	&  6~~~~	8	&  5~~~~	6	&  2~~~~	3	&  1~~~~	2\\
    2	&  6~~~~	7	&  5~~~~	6	&  1~~~~	2	&  0~~~~	1\\
    3	&  8~~~~	9	&  6~~~~	7	&  2~~~~	2	&  2~~~~	2\\
    4	&  8~~~~	9	&  5~~~~	6	&  2~~~~	2	&  1~~~~	2\\
    5	&  9~~~~	8	&  5~~~~	6	&  2~~~~	3	&  1~~~~	2\\
    6	&  6~~~~	8	&  3~~~~	5	&  2~~~~	2	&  1~~~~	2\\
    7	&  5~~~~	6	&  3~~~~	5	&  2~~~~	2	&  1~~~~	2\\
    8	&  4~~~~	7	&  2~~~~	5	&  1~~~~	2	&  1~~~~	2\\
    9	&  5~~~~	9	&  5~~~~	6	&  0~~~~	1	&  0~~~~	1\\
    10	&  7~~~~	10	&  4~~~~	6	&  1~~~~	2	&  0~~~~	1\\
    11	&  6~~~~	9	&  5~~~~	7	&  0~~~~	1	&  0~~~~	1\\
    12	&  7~~~~	6	&  5~~~~	6	&  3~~~~	2	&  1~~~~	2\\
    13	&  6~~~~	5	&  3~~~~	3	&  1~~~~	2	&  1~~~~	2\\
    14	&  3~~~~	7	&  2~~~~	5	&  1~~~~	2	&  1~~~~	2\\
    15	&  6~~~~	6	&  3~~~~	5	&  2~~~~	2	&  1~~~~	2\\
    16	&  6~~~~	6	&  6~~~~	6	&  1~~~~	2	&  1~~~~	2\\
    17	&  6~~~~	6	&  5~~~~	5	&  1~~~~	2	&  1~~~~	2\\
    18	&  4~~~~	6	&  3~~~~	4	&  0~~~~	1	&  0~~~~	1\\
    19	&  8~~~~	8	&  3~~~~	4	&  2~~~~	2	&  0~~~~	1\\
    20	&  6~~~~	9	&  4~~~~	4	&  2~~~~	3	&  1~~~~	2\\
    21	&  7~~~~	7	&  3~~~~	4	&  2~~~~	2	&  0~~~~	1\\
    22	&  5~~~~	5	&  5~~~~	5	&  2~~~~	2	&  2~~~~	2\\
    23	&  7~~~~	6	&  6~~~~	6	&  4~~~~	3	&  2~~~~	2\\
    24	&  4~~~~	5	&  5~~~~	5	&  0~~~~	1	&  0~~~~	1\\
    25	&  6~~~~	5	&  6~~~~	6	&  4~~~~	5	&  3~~~~	4\\
    26	&  7~~~~	6	&  5~~~~	4	&  4~~~~	4	&  2~~~~	3\\
    27	&  6~~~~	8	&  4~~~~	7	&  1~~~~	2	&  0~~~~	1\\
    28	&  7~~~~	5	&  6~~~~	6	&  3~~~~	4	&  2~~~~	3\\
    \hline
    $Average$	&  6.11 7.0 &	4.36 5.36 & 1.71  2.25 &	0.93 1.82\\
    \hline\hline
    \end{tabular}
    \end{table}

    Moreover, it is evident that when all the nodes in a WDM network are MC nodes, none of the shortest path trees constructed by the Dijkstra or the DijkstraPro algorithm will have any MIB nodes and their link stress will always be 1. So, it is obvious that when the ratio of MC nodes in the network is very high the improvement to be gained by using the DijkstraPro algorithm is not significant. But when the MC nodes are very sparse its performance is much better than the traditional Dijkstra algorithm, not only in terms of the number of MIB nodes but also in terms of the link stress. This justifies our introduction of the DijkstraPro algorithm in the shortest path tree construction step for the implementation of our proposed multicast routing algorithm.
	
    \subsection{Performance of the Avoidance of MIB Nodes Based Multicast Routing Algorithm}
    \label{sec:Performance of Avoidance of MIB Nodes based Multicast Routing Algorithm}
    There is no mention in the literature for the Reroute-to-Any~\cite{xjzhang2000} algorithm of a technique to determine which branch of a MIB node should be cut, and which algorithm should be used to reconnect the cut destinations. In our simulation an arbitrary branch is assumed to be kept and a Member-Only-like~\cite{xjzhang2000} reconnection method is employed.

    To evaluate the performance of the proposed multicast routing algorithm based on avoidance of MIB nodes (MIBPro/MIBPro2), the following four metrics are used to measure the quality of the multicast light-trees built for a multicast session:
    \begin{itemize}
      \item \emph{link stress}
      \item \emph{average end-to-end delay}
      \item \emph{maximum end-to-end delay}
      \item \emph{total cost}
    \end{itemize}
    In addition, each multicast session has only a single source. Each network node is selected as the source of a multicast session in turn. The destinations of a multicast group are distributed independently and uniformly through the network. For a given source and a given multicast group size, 100 random multicast sessions are generated. Hence, the result of each point in the simulation figures is the average of $100 \times |V|$ computations. In addition, Reroute-to-Source (R2S), Reroute-to-Any (R2A) and Member-Only (MO) are also implemented for comparison.

    \begin{figure*}
    \begin{center}
    $\begin{array}{c@{\hspace{1in}}c}
    \epsfxsize=2.21in \epsffile{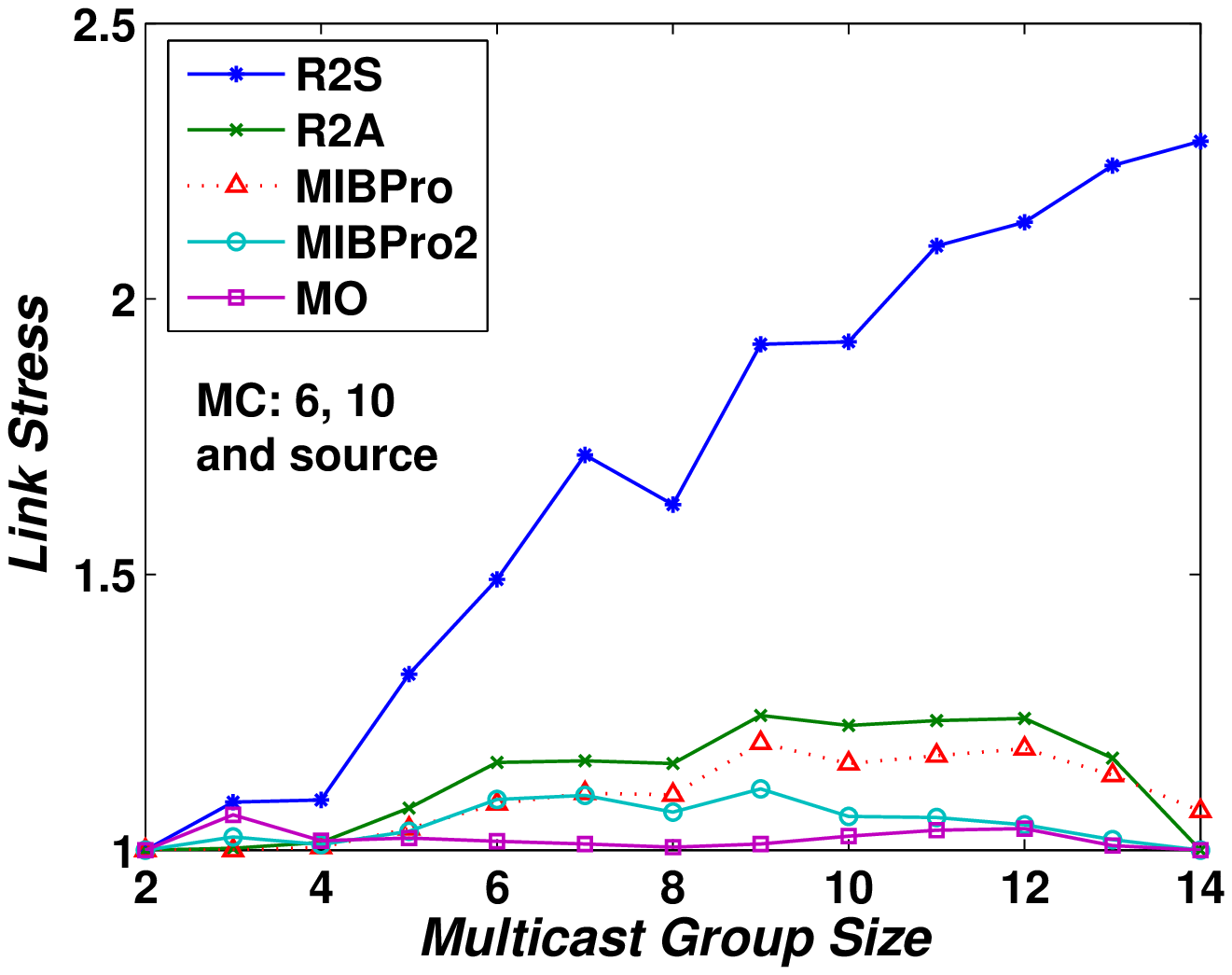} & \epsfxsize=2.21in \epsffile{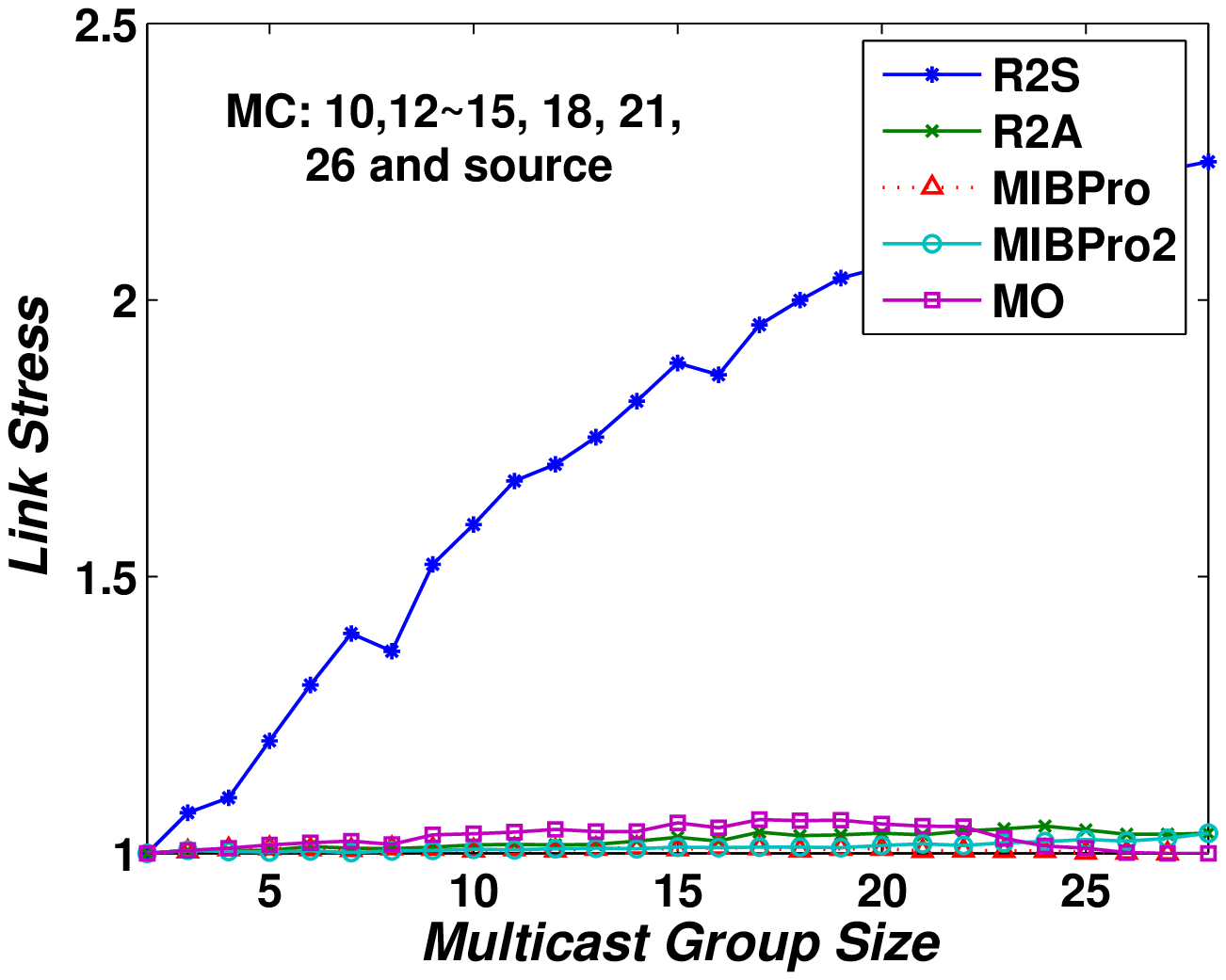} \\
    \mbox{\bf (a)} & \mbox{\bf (b)}
    \end{array}$
    \end{center}
    \caption{Link Stress vs. Multicast Group Size in (a)NSF Network (b)USA Longhaul Network}
    \label{link-stress-ndest}
    \end{figure*}

    \begin{figure*}
    \begin{center}
    $\begin{array}{c@{\hspace{1in}}c}
    \epsfxsize=2.21in \epsffile{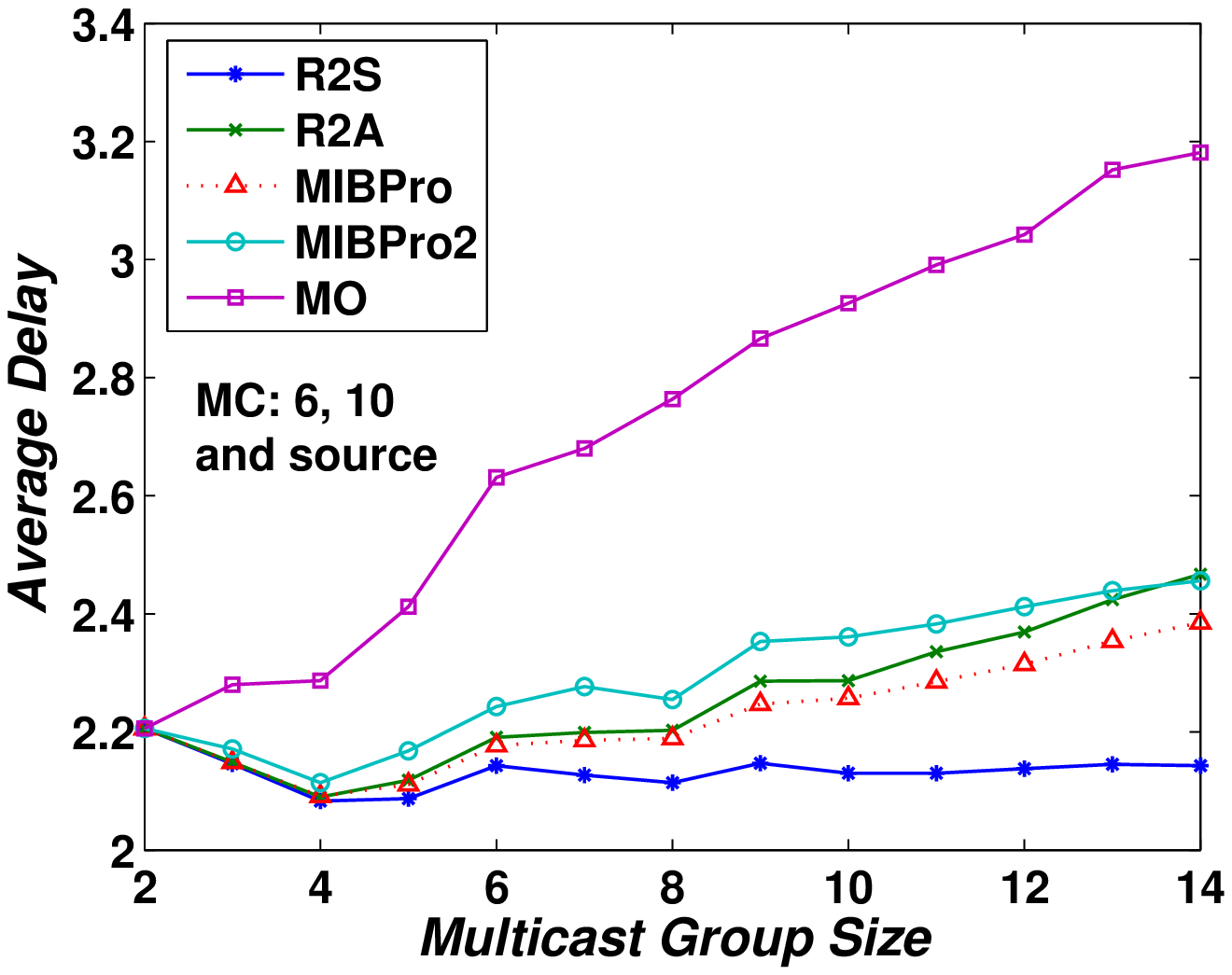} & \epsfxsize=2.21in \epsffile{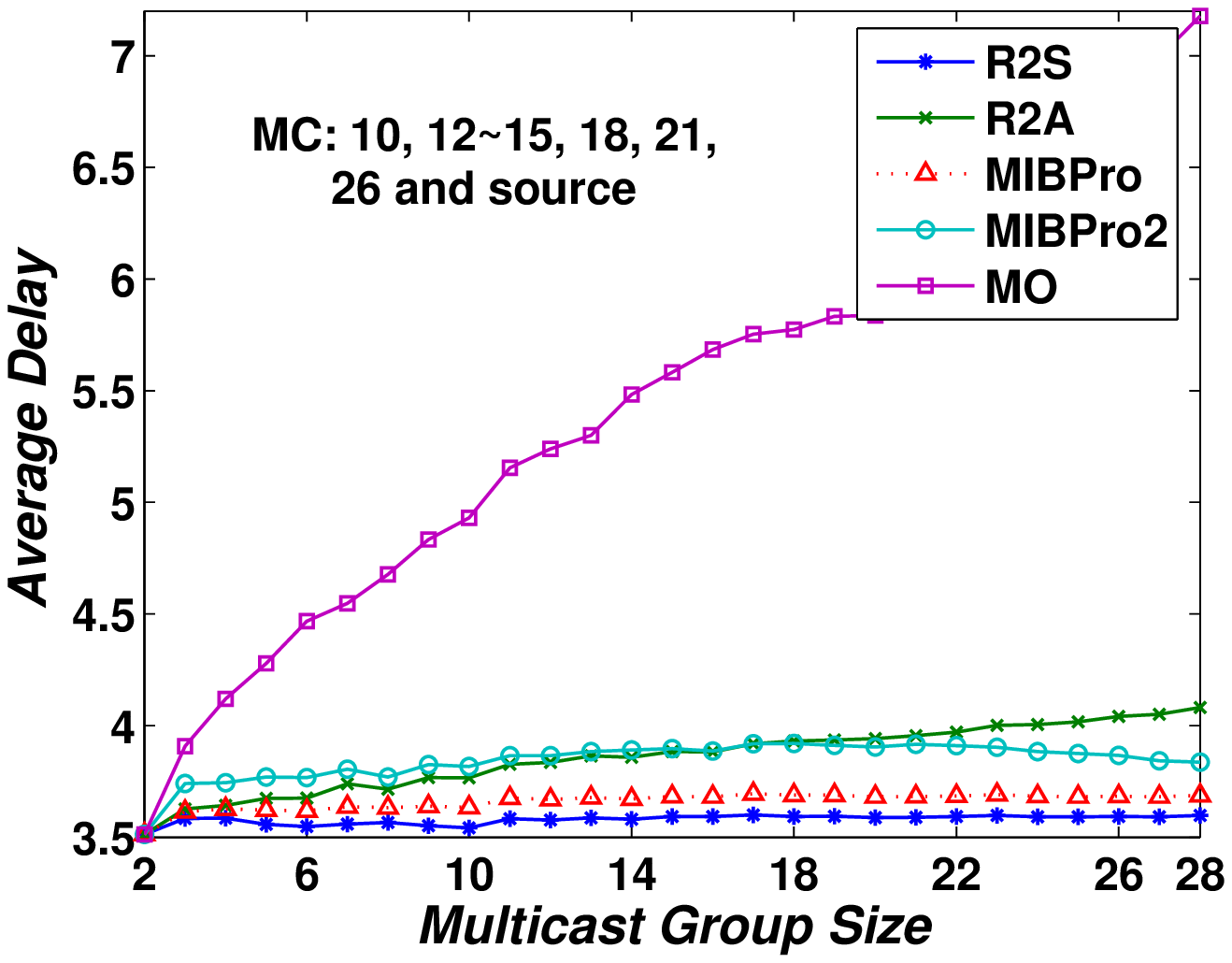} \\
    \mbox{\bf (a)} & \mbox{\bf (b)}
    \end{array}$
    \end{center}
    \caption{Average End-to-End Delay vs. Multicast Group Size in (a)NSF Network (b)USA Longhaul Network}
    \label{delay-ndest}
    \end{figure*}

    \begin{figure*}
    \begin{center}
    $\begin{array}{c@{\hspace{1in}}c}
    \epsfxsize=2.21in \epsffile{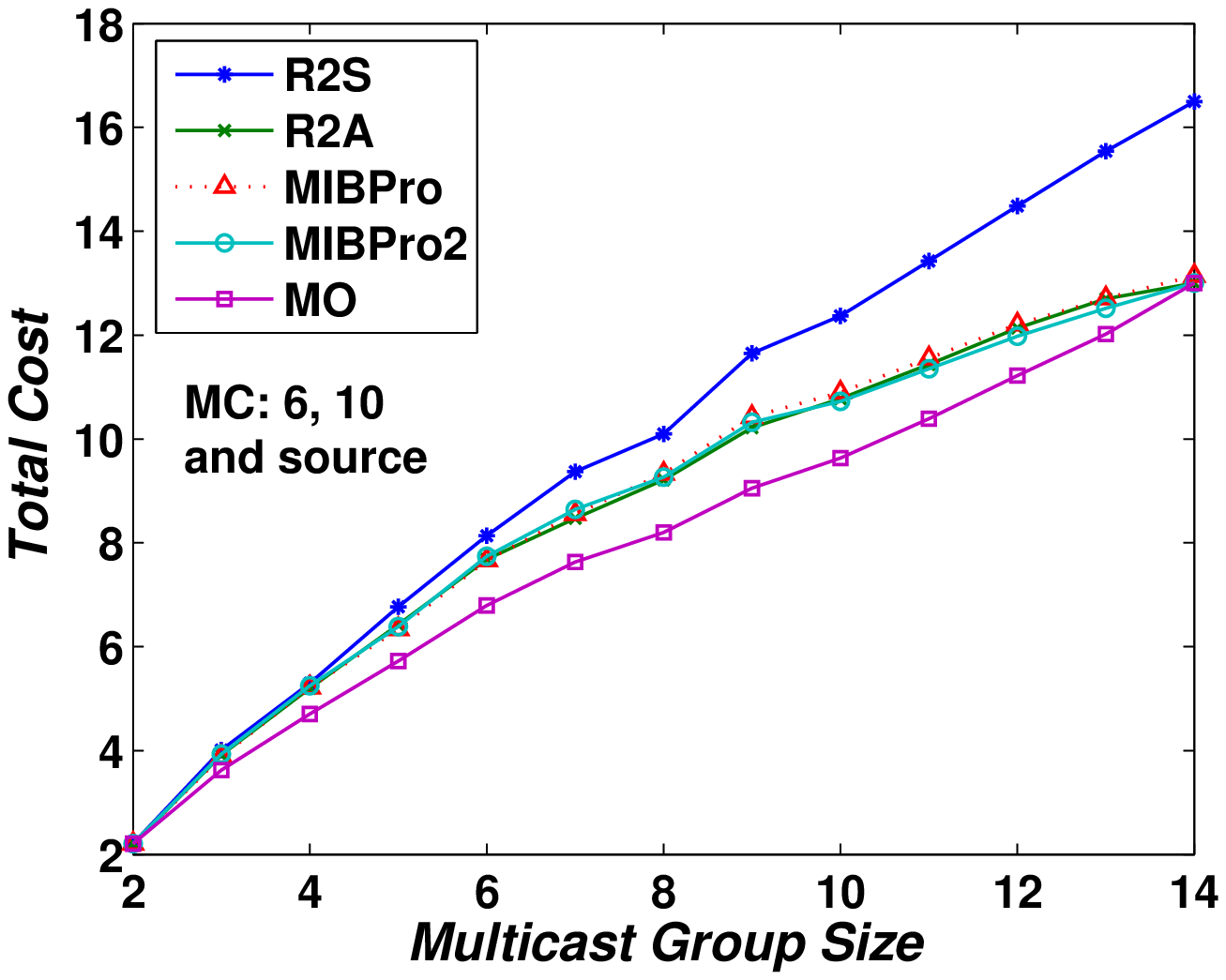} & \epsfxsize=2.21in \epsffile{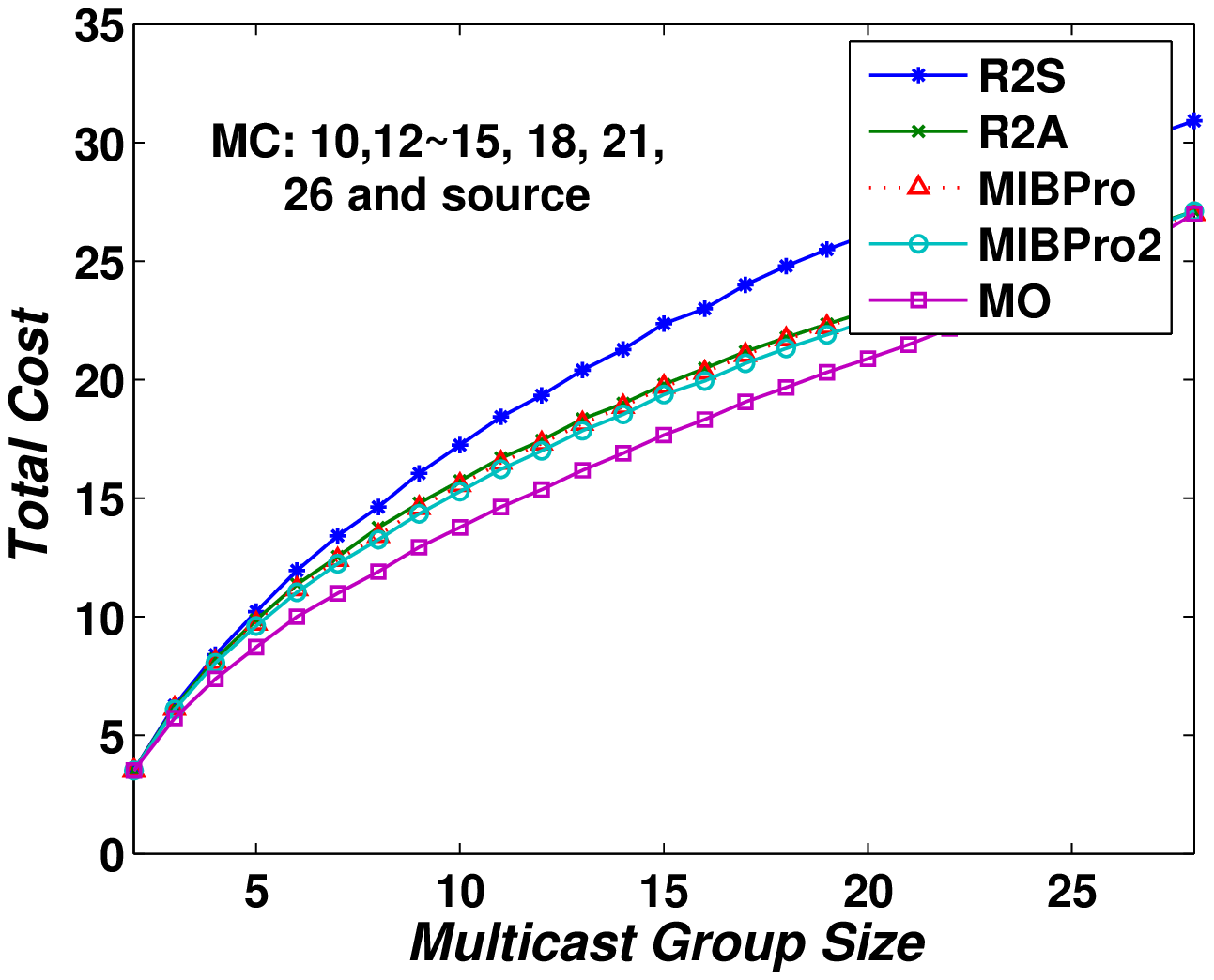} \\
    \mbox{\bf (a)} & \mbox{\bf (b)}
    \end{array}$
    \end{center}
    \caption{Total Cost vs. Multicast Group Size in (a)NSF Network (b)USA Longhaul Network}
    \label{cost-ndest}
    \end{figure*}

    \begin{figure*}
    \begin{center}
    $\begin{array}{c@{\hspace{1in}}c}
    \epsfxsize=2.21in \epsffile{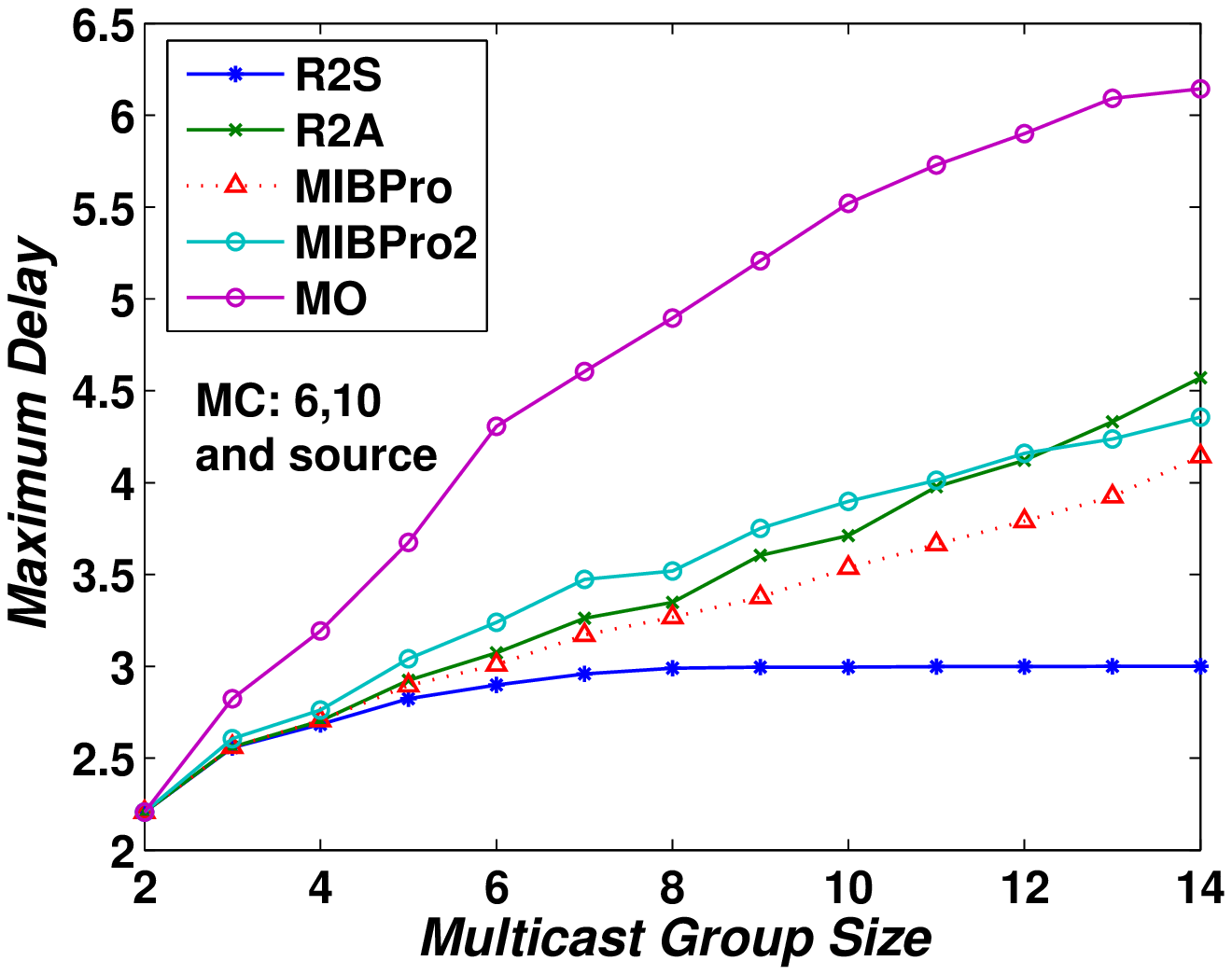} & \epsfxsize=2.21in \epsffile{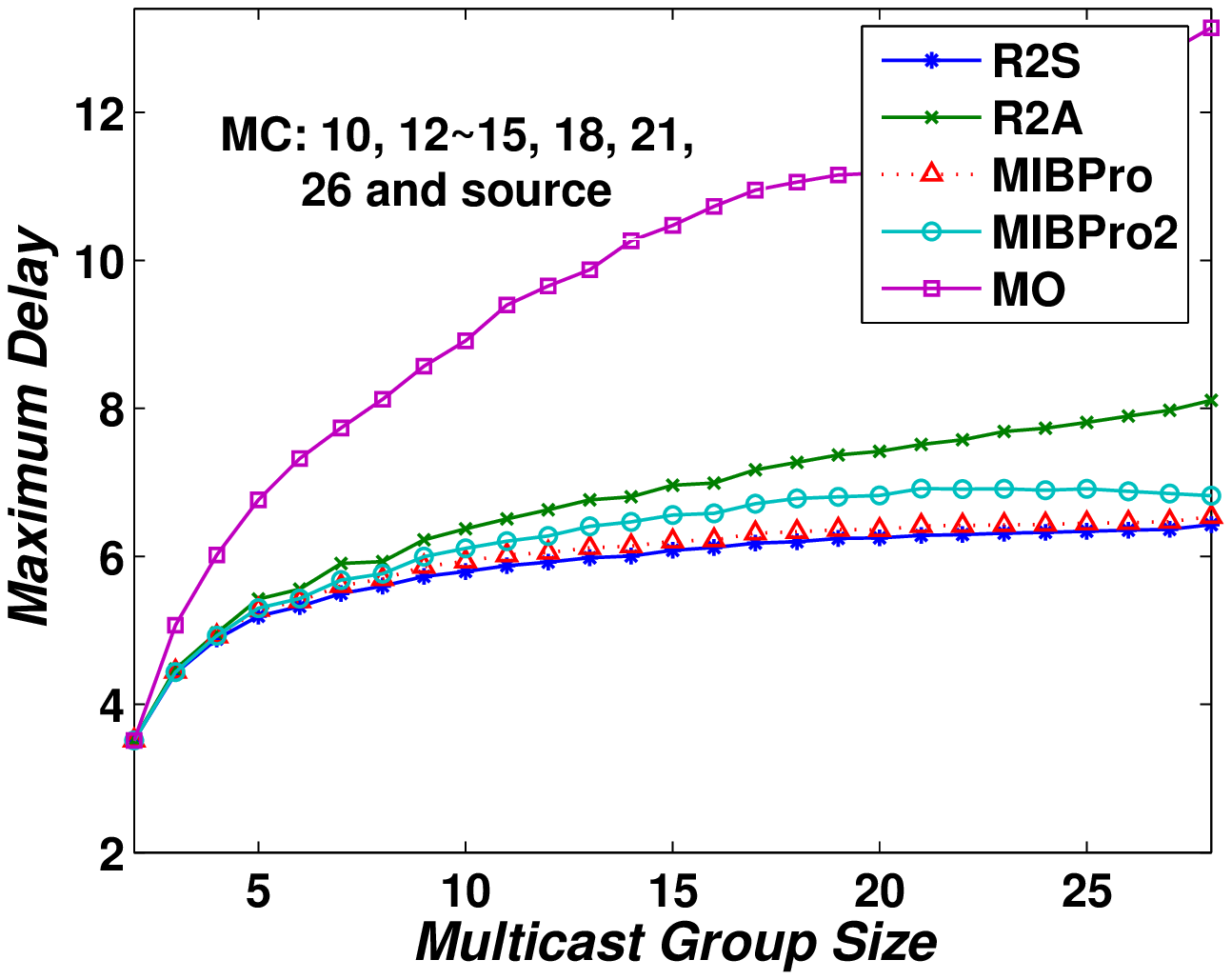} \\
    \mbox{\bf (a)} & \mbox{\bf (b)}
    \end{array}$
    \end{center}
    \caption{Maximum End-to-End Delay vs. Multicast Group Size in (a)NSF Network (b)USA Longhaul Network}
    \label{maxdelay-ndest}
    \end{figure*}

    \subsubsection{Effect of Group Size (Number of Multicast Members)}
    \label{sec:Effect of Group Size (Number of Multicast Members)}
    Here we study the performance of the proposed algorithm versus multicast group size. As mentioned in subsection~\ref{sec:Performance of DijkstraPro Algorithm}, nodes with high degree have a high probability of being MC nodes. To simplify the simulation in this part, we regard these nodes as MC nodes and only change the group size to evaluate the quality of light-trees built by MIBPro/MIBPro2 multicast routing algorithms.

    In the NSF network, nodes 6, 10, and the source are set as MC nodes. The simulation results in the NSF network are plotted in Figures~\ref{link-stress-ndest}-\ref{maxdelay-ndest}(b). As shown in Figure~\ref{link-stress-ndest}(a), we can see that when the group size is above four, MIBPro achieves better link stress than R2A. The link stress of MIBPro2 is also much smaller than MIBPro. Figures~\ref{delay-ndest}(a) and \ref{maxdelay-ndest}(a) show that the average end-to-end delay and maximum end-to-end delay of MIBPro is only second to the optimal result of R2S. As multicast group size grows the improvement of end-to-end delay returned by MIBPro compared to R2A becomes more and more significant. Moreover, while the total costs of R2A, MIBPro and MIBPro2 are almost the same, R2S results in the highest and MO results in the lowest total cost.

    In the USA Longhaul network, nodes 10, 12$\sim$15, 18, 21, 26, and the source are set as MC nodes. Figures~\ref{link-stress-ndest}-\ref{maxdelay-ndest}(b) have compared the performance of those five algorithms in this topology. The link stress of the five algorithms are almost the same and very near to 1 according to Figure~\ref{link-stress-ndest}(b). This is because the ratio of MC nodes is very high (32\%) in this configuration. The end-to-end delay for the MIBPro algorithm is very close to the optimum (R2S). To our surprise, MIBPro obtains almost the same maximum end-to-end delay as R2S. From the point of view of the total cost, R2A, MIBPro and MIBPro2 return the same value, which is the same outcome as the NSF network example.

    In both topologies the performance of R2S in terms of link stress and total cost is always the worst, while its performance in end-to-end delay is the best. Conversely the MO algorithm can achieve very good link stress and total cost, while its end-to-end delay is too large.

    From the simulation results above it can be seen that the MIBPro algorithm can provide nearly the same or even slightly better link stress than R2A. Its reduction in average and maximum end-to-end delay compared to R2A becomes more obvious when the group size is large. This is because the MC node priority mechanism, node adoption and distance based reconnection do not affect the result when the group size is too small. Only when there are enough destinations can these strategies work well. Overall, however, the MIBPro algorithm achieves a good tradeoff between link stress and end-to-end delay.

    \subsubsection{Effect of Splitting Capability (Number of MC nodes)}
    \label{sec: Effect of Group Size (Number of Multicast Members)}
\begin{figure*}
        \begin{center}
        $\begin{array}{c@{\hspace{1in}}c}
        \epsfxsize=2.21in \epsffile{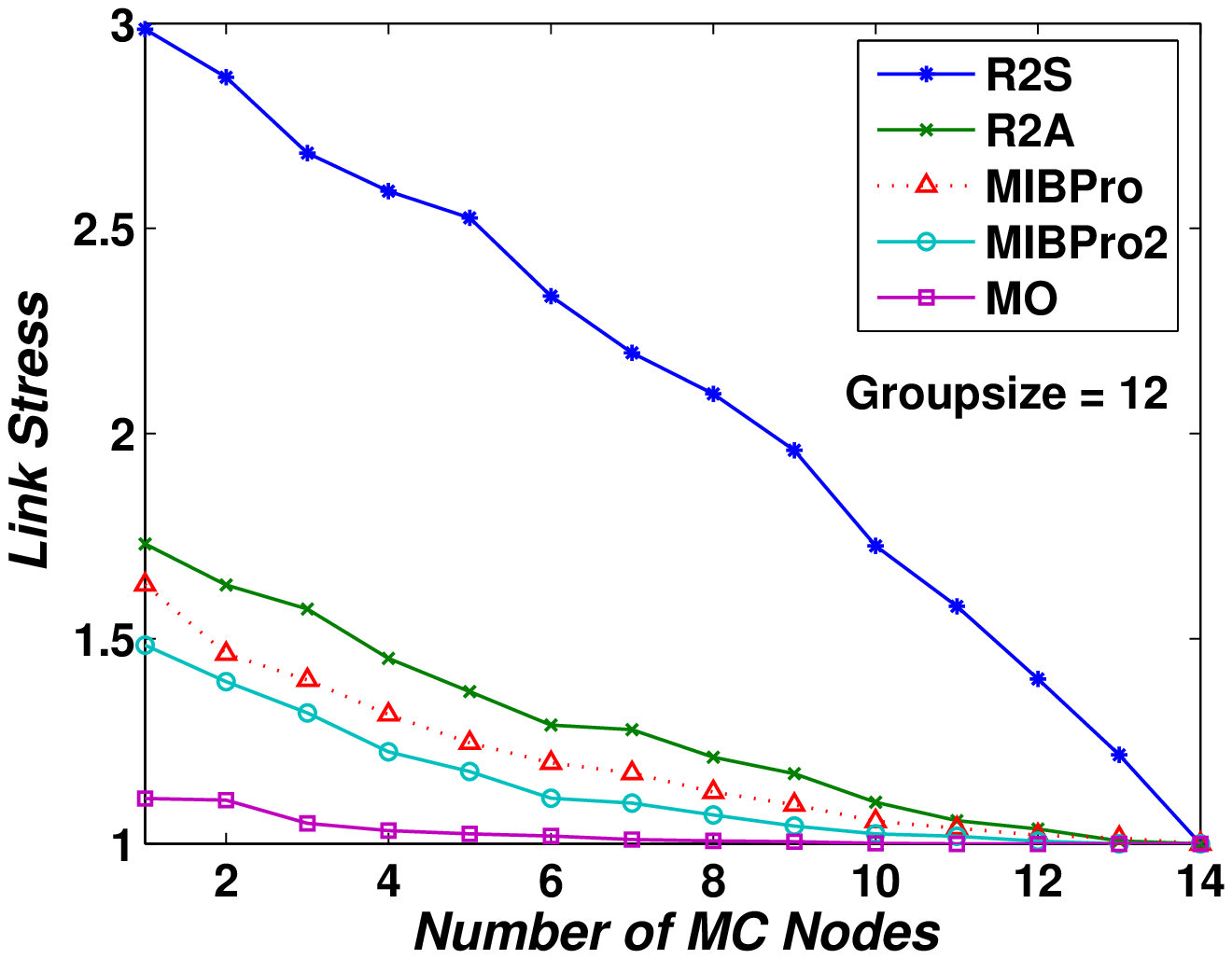} & \epsfxsize=2.21in \epsffile{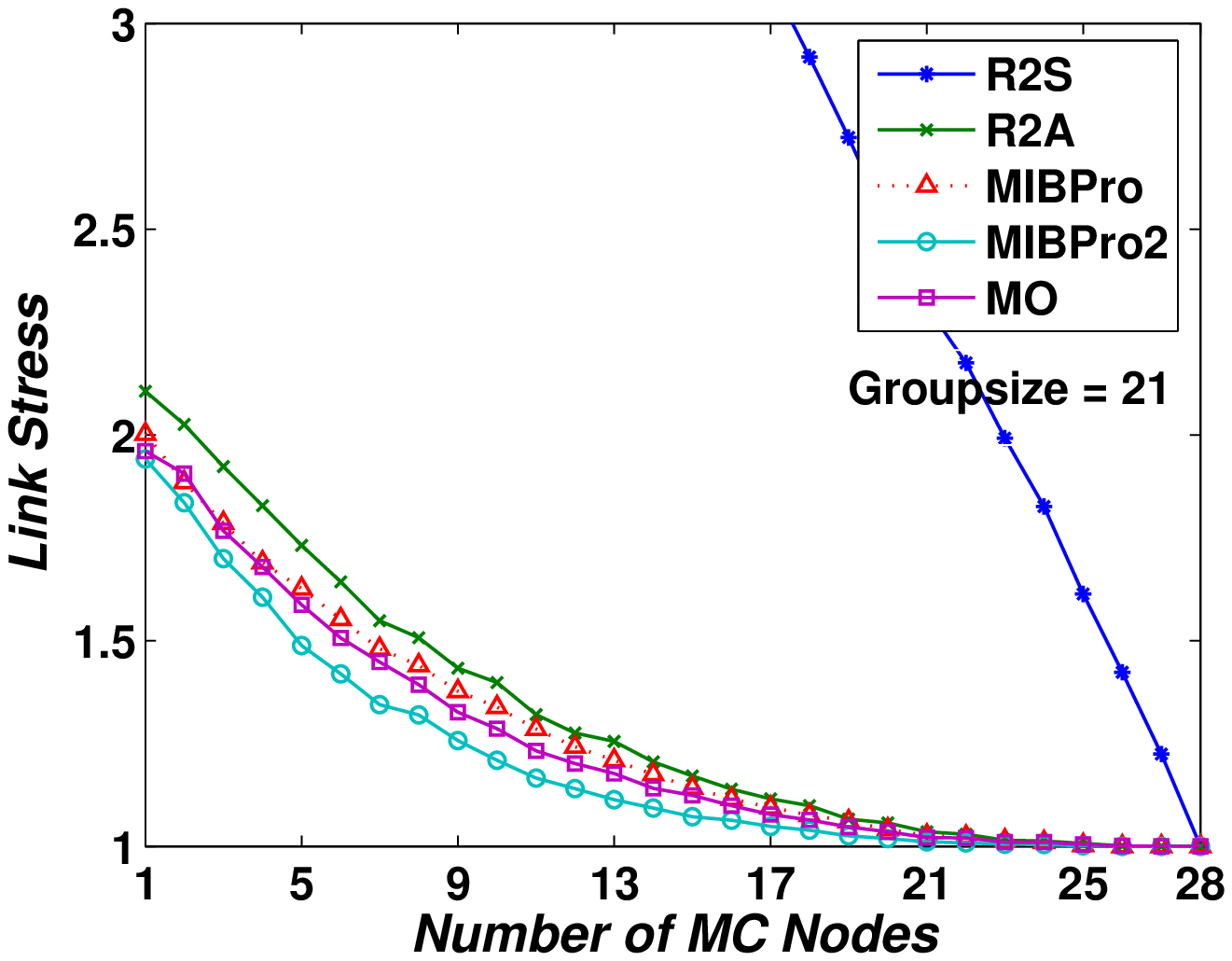} \\
        \mbox{\bf (a)} & \mbox{\bf (b)}
        \end{array}$
        \end{center}
        \caption{Link Stress against the Number of MC Nodes in (a)NSF Network (b)USA Longhaul Network}
        \label{link-stress-mc}
        \end{figure*}

\begin{figure*}
        \begin{center}
        $\begin{array}{c@{\hspace{1in}}c}
        \epsfxsize=2.21in \epsffile{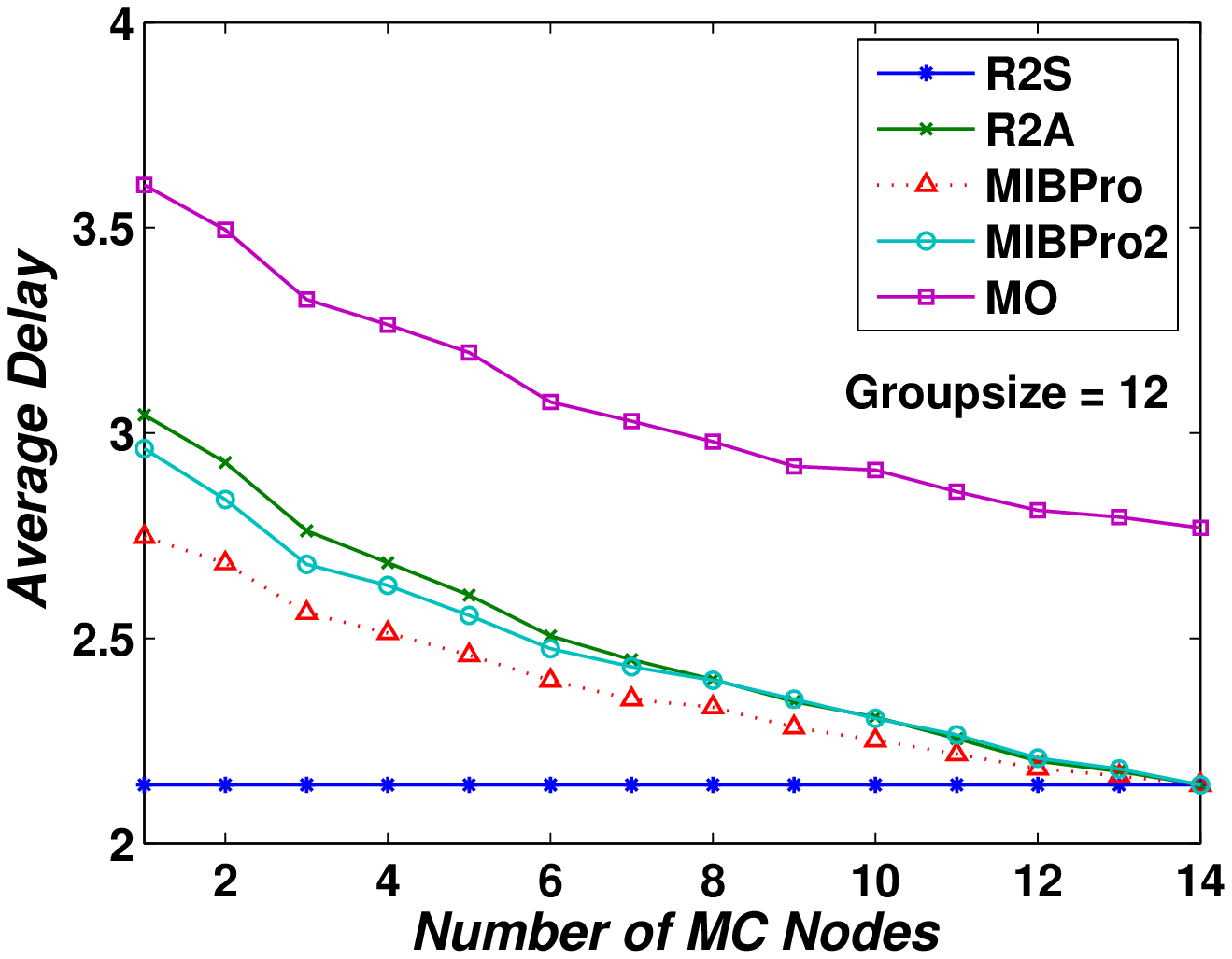} & \epsfxsize=2.21in \epsffile{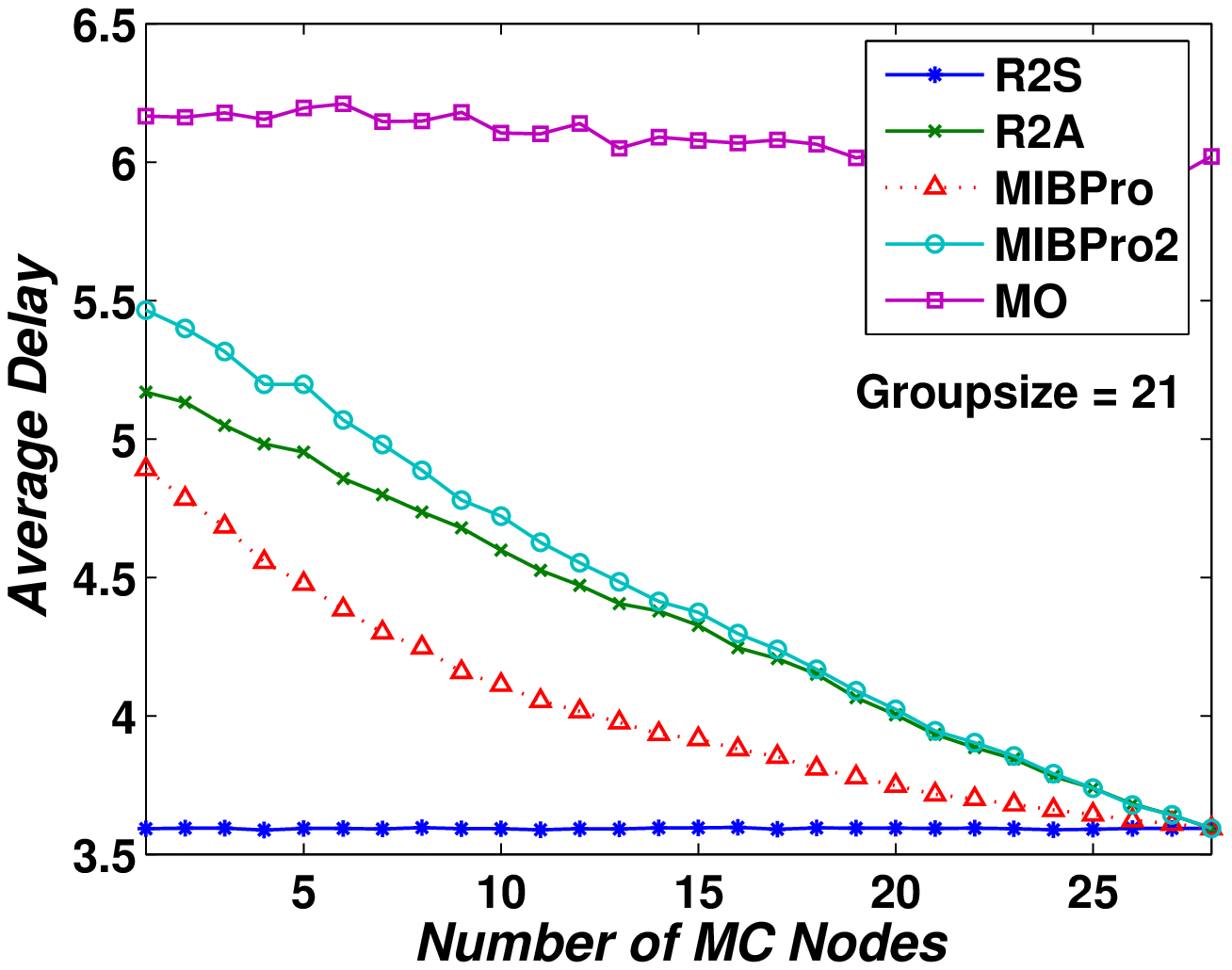} \\
        \mbox{\bf (a)} & \mbox{\bf (b)}
        \end{array}$
        \end{center}
        \caption{Average End-to-End Delay against the Number of MC Nodes in (a)NSF Network (b)USA Longhaul Network}
        \label{delay-mc}
        \end{figure*}

\begin{figure*}
        \begin{center}
        $\begin{array}{c@{\hspace{1in}}c}
        \epsfxsize=2.21in \epsffile{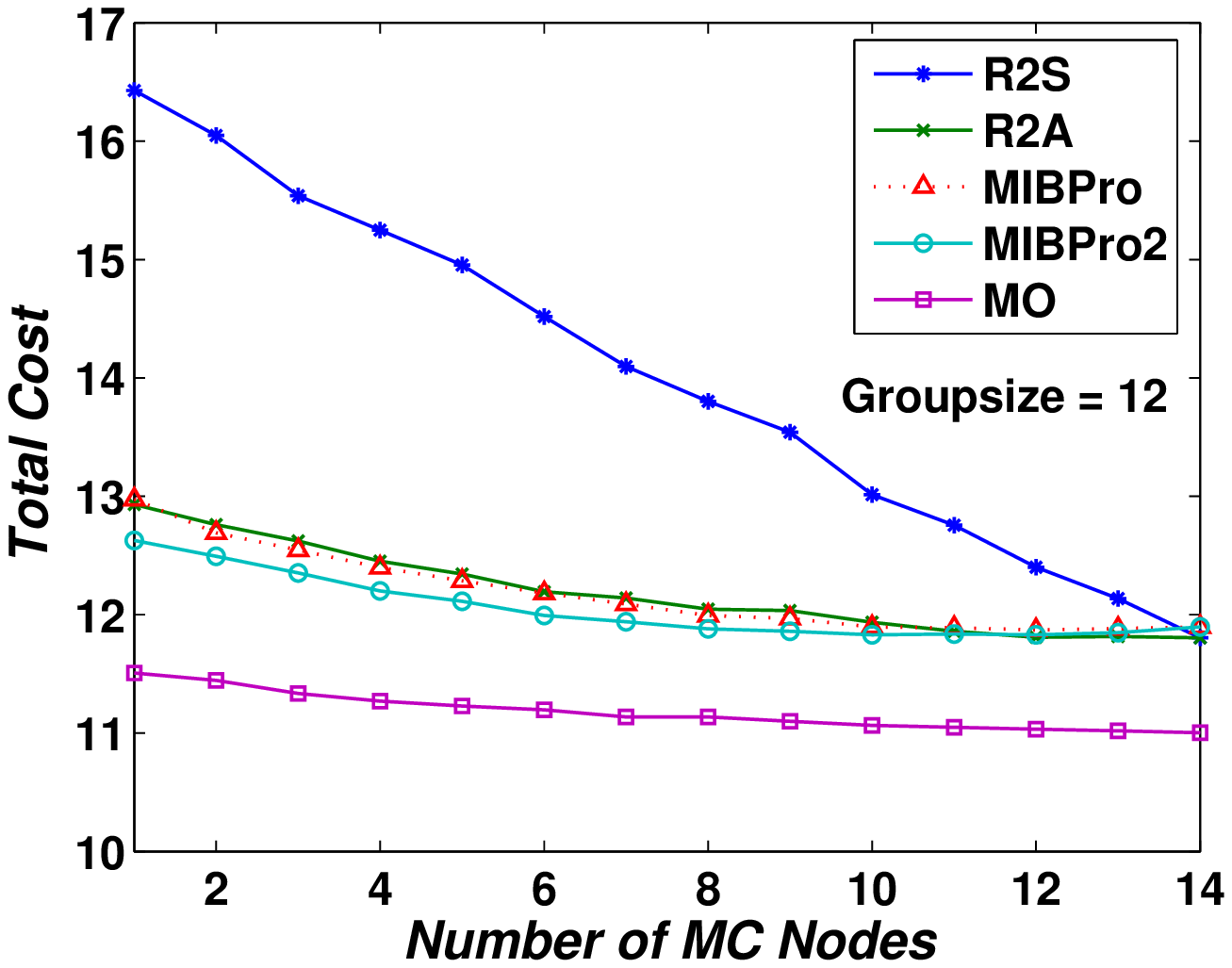} & \epsfxsize=2.21in \epsffile{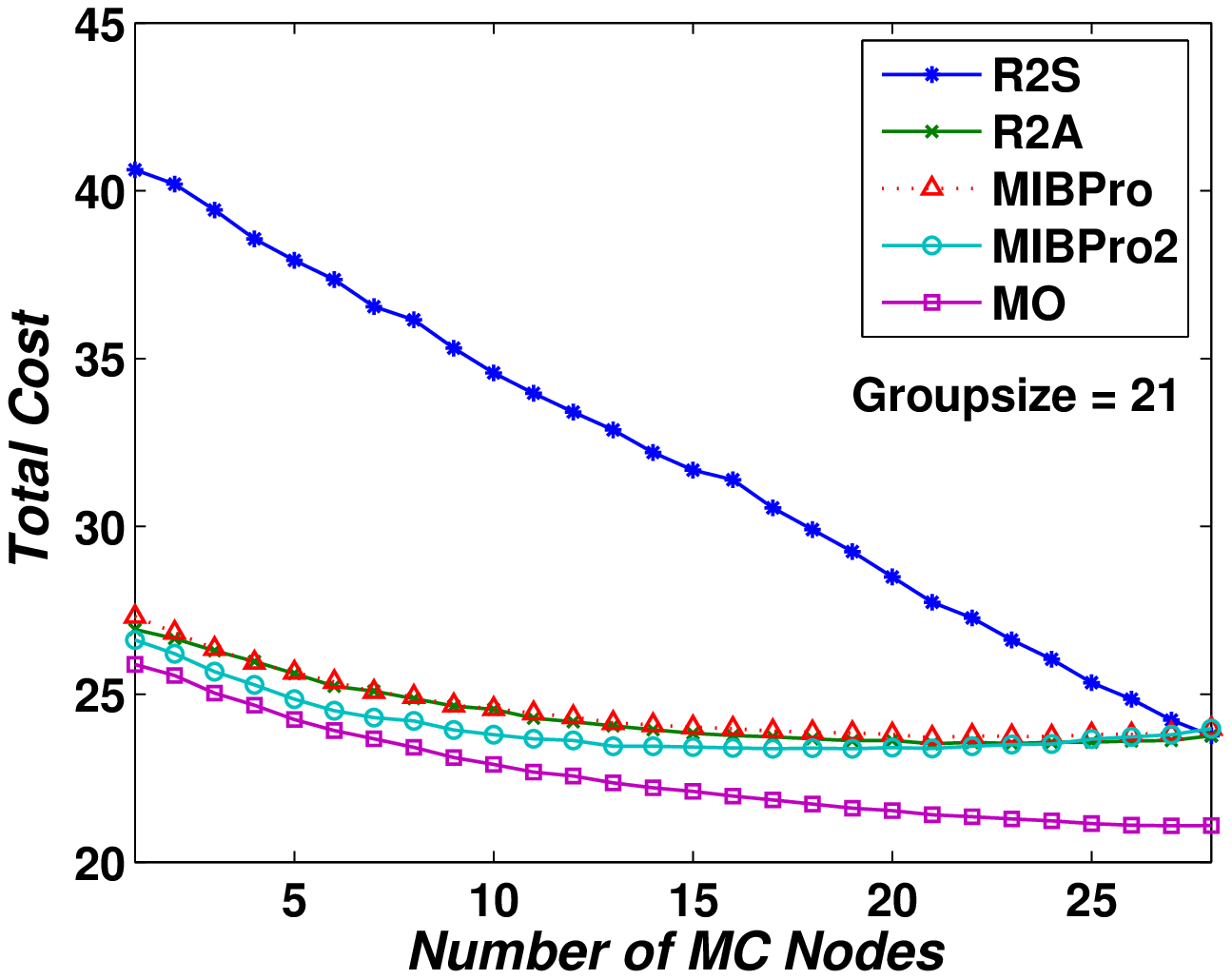} \\
        \mbox{\bf (a)} & \mbox{\bf (b)}
        \end{array}$
        \end{center}
        \caption{Total Cost against the Number of MC Nodes in (a)NSF Network (b)USA Longhaul Network}
        \label{cost-mc}
        \end{figure*}

\begin{figure*}
        \begin{center}
        $\begin{array}{c@{\hspace{1in}}c}
        \epsfxsize=2.21in \epsffile{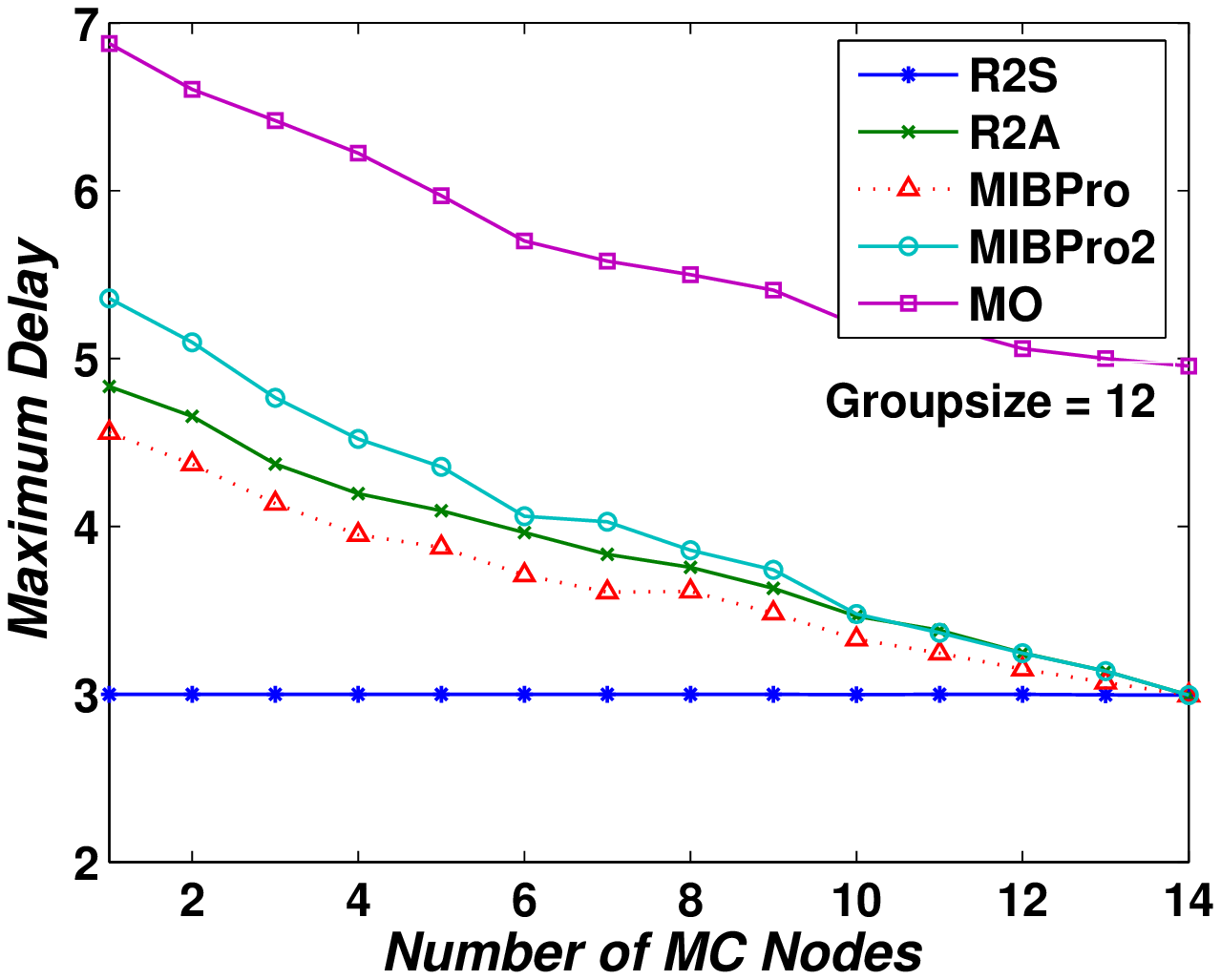} & \epsfxsize=2.21in \epsffile{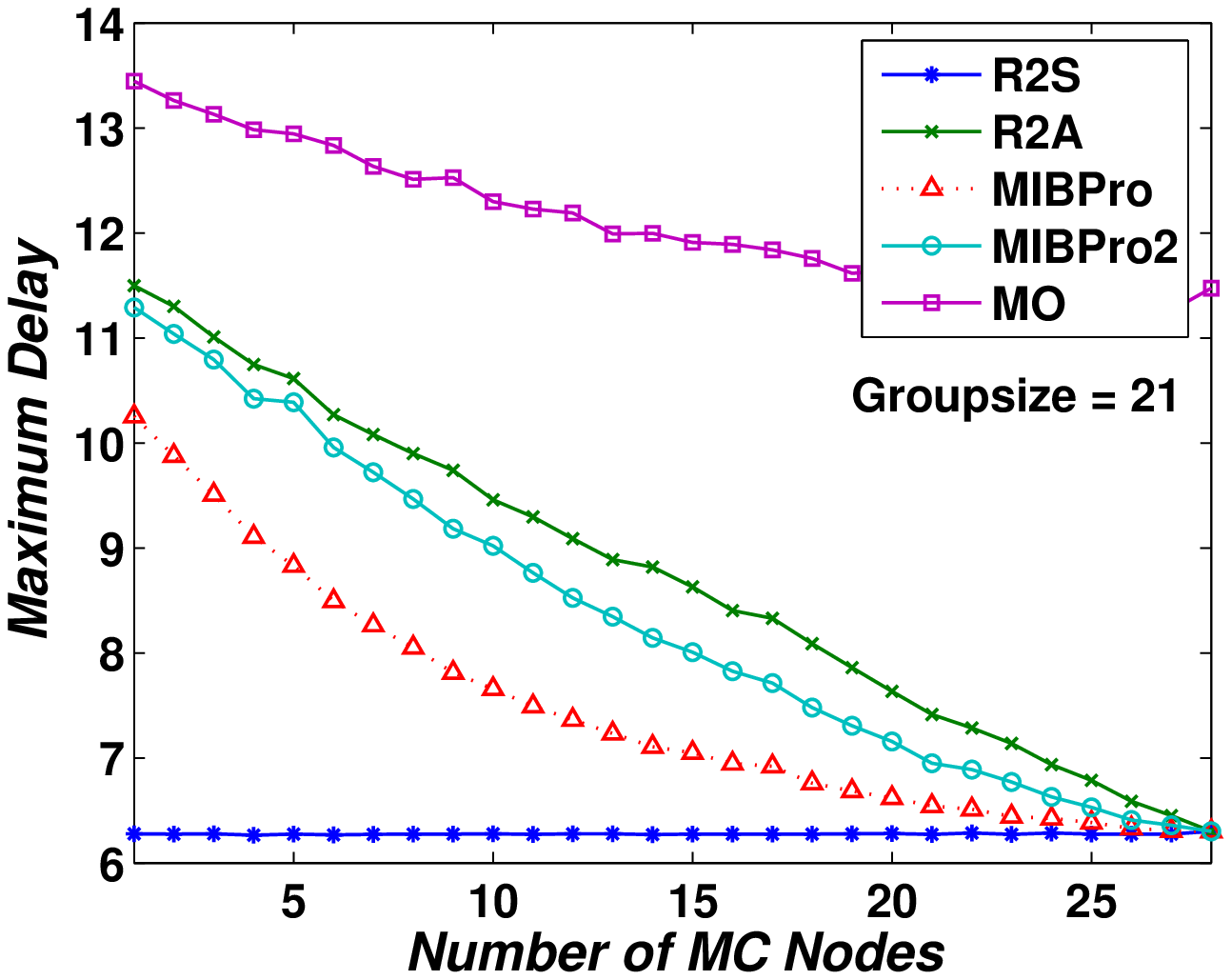} \\
        \mbox{\bf (a)} & \mbox{\bf (b)}
        \end{array}$
        \end{center}
        \caption{Maximum End-to-End Delay against the Number of MC Nodes in (a)NSF Network (b)USA Longhaul Network}
        \label{maxdelay-mc}
        \end{figure*}

    The performance when the number of MC nodes varies have also been studied. According to the results of the previous section, MIBPro is more advantageous when the multicast group size is large. Thus, the multicast group size is set at a large value while only the number of MC nodes is changed in the simulation of this part. The MC nodes are assumed to be independently and uniformly distributed in the topology. The multicast group size is set to 12 in the 14 nodes NSF network and set to 21 in the 28 nodes USA Longhaul network. The numeric results are plotted in Figures~\ref{link-stress-mc}-\ref{maxdelay-mc}. According to these figures, when MC nodes are sparse, (1) MIBPro achieves much better performance in terms of link stress, average end-to-end delay and maximum end-to-end delay relative to R2A while producing the same cost as R2S. (2) MIBPro2 results in both lower link stress and total cost than R2A. Its link stress is even better than MO in the Longhaul network. However, its end-to-end delay is either better or worse than R2A.

    These results indicate that our proposed MIBPro algorithm works well in the case of sparse light splitting. When the ratio of MC nodes is large, there are fewer MIB nodes in the shortest path tree and as a result MIBPro's advantage is less significant.
   % \begin{figure}
%    \begin{center}
%    $\begin{array}{c@{\hspace{1in}}c}
%    \multicolumn{1}{l}{\mbox{\bf (a)}} &
%    	\multicolumn{1}{l}{\mbox{\bf (b)}} \\ [-0.53cm]
%  \epsfxsize=2in
%    \epsffile{link-stress-nsf-dest4.eps} &
%  \epsfxsize=2in
%    	\epsffile{link-stress-nsf-dest12.eps} \\ [0.4cm]
%    \mbox{\bf (aa)} & \mbox{\bf (bb)}
%    \end{array}$
%   \end{center}
%    \caption{The caption for Figure \protect\ref{figtest-fig}}
%    \label{figtest-fig}
%    \end{figure}

    \section{Conclusion}
    \label{sec:Conclusion}
    Due to physical constraints multicast routing in WDM networks with sparse light splitting is not easy. Many multicast routing algorithms have been proposed to find the Steiner based light-tree in WDM networks with the minimum cost, but it has been proved that this problem is NP-hard. Applications with QoS requirements are becoming more and more popular in the Internet. The bandwidth (or number of wavelengths supported per fiber link in WDM networks) and the end-to-end delay are two important parameters for QoS. Hence, a multicast routing algorithm based on avoidance of MIB nodes is presented for traffic with QoS requirements in WDM networks in order to decrease the link stress and the end-to-end delay. The algorithm retains the parts of the shortest path tree which result in optimal end-to-end delay for at least some multicast members. In order to reduce the number of MIB nodes and the link stress in the construction of the shortest path tree step the DijkstraPro algorithm is presented, where a higher priority is assigned to MC \emph{candidate} nodes and node adoption is performed between the \emph{candidate} nodes at the same \emph{level}. To keep one branch of MIB nodes in the shortest path tree, critical articulation and deepest branch heuristics are introduced. Finally, the distance-based light-tree reconnection algorithm is developed to rejoin the multicast light-forest. The first part of the simulation in Section~\ref{sec:Performance Evaluation and Simulation} shows that the DijkstraPro algorithm is a better tool for shortest path tree construction in all-optical networks than the traditional Dijkstra algorithm. It can really reduce the number of MIB nodes and the link stress of the shortest path tree. Moreover, the second part of the simulation proves that	the proposed MIBPro algorithm yields good performance in terms of link stress when MC nodes are very sparse. In addition, when the group size is large enough it is able to improve the average and maximum end-to-end delay dramatically giving a result very close to the optimal Reroute-to-Source algorithm solution~\cite{xjzhang2000}. To sum up, the proposed algorithm is a good tradeoff between link stress and end-to-end delay for multicast routing in sparse light splitting WDM networks.


\begin{thebibliography}{}

\bibitem{rMalli1998}
R. Malli, Xijun Zhang, Chunming Qiao. Benefit of Multicasting in all-optical networks. SPIE Proceeding on All-Optical Networking, 2531:209-220, 1998.

\bibitem{lhSahasrabuddhe1999}
L. H. Sahasrabuddhe, B. Mukherjee. Light-tree: Optical multicasting for improved performance in wavelength-routed networks. IEEE Communication. Magazine, 37: 67-73, 1999.

\bibitem{bMukherjee2000}
Biswanath Mukherjee. WDM optical communication networks: Progress and challenges. IEEE Journal on Selected Areas in Communications, 18(10):1810-1824, 2000.

\bibitem{xjzhang2000}
Xijun Zhang, John Wei, Chunming Qiao. Constrained Multicast Routing in WDM Networks with Sparse Light Splitting. Journal of Lightware Technology, 18(12):1917-1927, 2000.

\bibitem{nSreenath2001Photonic}
N. Sreenath, K. Satheesh, G. Mohan, C. Siva Ram Murthy. Virtual Source Based Multicast Routing in WDM Optical Networks. Photonic Network Communications, 3(3): 213-226, 2001.

\bibitem{nSreenath2001High}
N. Sreenath, K. Satheesh, G. Mohan, C. Siva Ram Murthy. Virtual Source Based Multicast Routing in WDM Networks with Sparse Light Splitting. IEEE Workshop on High Performance Switching and Routing'01, p141-145, 2001.

\bibitem{aHamad2006}
Ashraf Hamad, Tao Wu, Ahmed E. Kamal, Arun K. Somani. On multicasting in wavelength-routing mesh networks. Computer Networks, 50: 3105-3164, 2006.

\bibitem{hTakahashi1980}
H. Takahashi, A. Matsuyama. An approximate solution for the Steiner problem in graphs. Math. Japonica, 24(6): 573-577, 1980.

\bibitem{aZsigri2003}
Aniko Zsigri, Alexandre Guitton, Miklos Molnar. Construction of light-trees for WDM multicasting under splitting capability constraints. International Conference on Telecomunication'03, p171-175, 2003.

\bibitem{sgYani2003}
Shuguang Yan, Jitender S. Deogun, Maher Ali. Network Routing in sparse splitting optical networks with multicast traffic. Computer Network, 41: 89-113, 2003.

\bibitem{csrMurthy2002}
C. Siva Ram Murthy, Mohan Gurusamy. WDM Optical Networks: concepts, design and algorithms. Prentice Hall, 2002.

\bibitem{mAli2000Cost}
Maher Ali, Jitender S. Deogun. Cost-Effective Implementation of Multicasting in Wavelength-Routed Networks. IEEE Journal of Lightwave Technology, 18(12): 1628-1638, 2000.

\bibitem{mAli2000Allocation}
Maher Ali, Jitender S. Deogun. Allocation of Splitting Nodes in All-Optical Wavelength-Routed Networks. Photonic Network Communication, 2(3): 247-265, 2000.

\bibitem{fZhou2008LCN}
Fen Zhou, Mikl\a'os Moln\a'ar, Bernard Cousin. Avoidance of Multicast Incapable Branching Nodes for Multicast Routing in WDM Networks. The 33$^{rd}$ IEEE International Conference on Local Computer Network, pp336-344, Montr\a'eal, Canada, October, 2008.

\bibitem{eDesurvire1991}
E. Desurvire. Erbium-Doped Fiber Amplifiers: Principles and Applications. Wiley, New York, 1991.

\bibitem{fZhou2008ICCS}
Fen Zhou, Mikl\a'os Moln\a'ar, Bernard Cousin. Distance Priority Based Multicast Routing in WDM Networks Considering Sparse Light Splitting. The 11$^{th}$ IEEE International Conference on Communication System, pp709-714, Guangzhou, China, November, 2008.

\bibitem{zgZhang2007}
Zhenghao Zhang, Yuanyuan Yang. On-Line Optimal Wavelength Assignment in WDM Networks with Shared Wavelength Converter Pool. IEEE Transaction On Networking, 15(1):234-245, 2007.

\bibitem{hLin2005}
Hwachun Lin, Shengwei Wang. Splitter Placement in All-Optical WDM Networks. In proceeding of IEEE GlobalCom, p306-310, 2005.

\bibitem{aBillah2007}
Abdur Billah, Bin Wang, Abdul A. S. Awwal. Topology based Placement of Multicast Capable Nodes for Supporting Efficient Multicast Communication in WDM Optical Networks. Photonic Network Communication, 14:35-47, 2007.


\end{thebibliography}
\end{document}